\documentclass[review]{elsarticle}
\usepackage{amsmath,amssymb}
\usepackage{lipsum}
\usepackage{bm}
\usepackage{graphicx}
\usepackage{graphics}
\usepackage{amsmath}
\usepackage{array}
\usepackage[caption=false]{subfig}
\usepackage{xcolor}
\usepackage{subfig}
\usepackage{hyperref}
\usepackage[capitalise]{cleveref}
\usepackage{textcomp}

\usepackage{epstopdf}
\usepackage{cases}
\usepackage{amssymb}
\usepackage{multirow}
\usepackage{siunitx}
\usepackage{cases}
\usepackage{tabularx}
\usepackage[top=1.1in, bottom=1.1in, left=1.0in, right=1.0in]{geometry}
\usepackage[utf8]{inputenc}
\usepackage{tikz}
\usepackage{comment}
\usepackage{lineno,hyperref}
\modulolinenumbers[5]
\usepackage{booktabs}

\begin{document}
\title{A novel transient heat conduction phenomenon to distinguish the hydrodynamic and (quasi) ballistic phonon transport}
\author[add1]{Chuang Zhang}
\ead{zhangcmzt@hust.edu.cn}
\author[add1]{Zhaoli Guo \corref{cor1}}
\ead{zlguo@hust.edu.cn}
\cortext[cor1]{Corresponding author}
\address[add1]{State Key Laboratory of Coal Combustion, School of Energy and Power Engineering, Huazhong University of Science and Technology, Wuhan 430074, China}
\date{\today}

\begin{abstract}

{\color{black}{Previous studies have predicted the failure of Fourier's law of thermal conduction due to the existence of wave like propagation of heat with finite propagation speed.}}
This non-Fourier thermal transport phenomenon can appear in both the hydrodynamic and (quasi) ballistic regimes.
Hence, it is not easy to clearly distinguish these two non-Fourier regimes only by this phenomenon.
In this work, the transient heat propagation in homogeneous thermal system is studied based on the phonon Boltzmann transport equation (BTE) under the Callaway model.
Given a quasi-one or quasi-two (three) dimensional simulation with homogeneous environment temperature, at initial moment, a heat source is added suddenly at the center with high temperature, then the heat propagates from the center to the outer.
Numerical results show that in quasi-two (three) dimensional simulations, the transient temperature will be lower than the lowest value of initial temperature in the hydrodynamic regime within a certain range of time and space.
This phenomenon appears only when the normal scattering dominates heat conduction.
Besides, it disappears in quasi-one dimensional simulations.
Similar phenomenon is also observed in thermal systems with time varying heat source.
This novel transient heat propagation phenomenon of hydrodynamic phonon transport distinguishes it well from (quasi) ballistic phonon transport.

\end{abstract}
\begin{keyword}
Hydrodynamic/(Quasi) ballistic phonon transport \sep Multiscale transient heat propagation \sep Boltzmann transport equation
\end{keyword}
\maketitle

\section{INTRODUCTION}

The transient heat conduction of bulk materials at room temperature is usually described by the Fourier's law of thermal conduction.
In the absence of time-varying heat source or oscillatory boundaries, the temperature always exponentially decays with time due to involvement of first-order time-derivative in the diffusion equation.
However, when the characteristic length/time of the thermal system is comparable to the phonon mean free path/relaxation time~\cite{ChenG05Oxford,second_sound_ge2020,PhysRevB.101.075303,PhysRevLett.59.1962,PhysRevB.72.125413}, or when the momentum-conserved normal (N) scattering dominates heat conduction~\cite{PhysRevLett.16.789,cepellotti_phonon_2015,lee_hydrodynamic_2015,huberman_observation_2019}, the Fourier's law may be broken~\cite{esee8c146,leesangyeopch1,PhysRevB.100.085203,PhysRevB.102.104310} and the wave like propagation of heat with finite propagation speed~\cite{RevModPhysJoseph89,cattaneo1948sulla,tzou1995a,luo2019,nie2020thermal,PhysRevLett.125.265901} may appear.

In the past decades, many theoretical, numerical and experimental studies have been conducted on the wave like propagation of heat~\cite{RevModPhysJoseph89,WangMr15application,leesangyeopch1} in the non-Fourier regime.
One of the underlying physical mechanisms of wave like propagation of heat is the hydrodynamic phonon transport~\cite{ENZ1968114,Dreyer1993,huberman_observation_2019,PhysRevLett_secondNaF,PhysRevB_SECOND_SOUND,PhysRevLett.28.1461}, which emerges when the N scattering frequently happens and the momentum-destroying resistive (R) scattering rarely happens.
In order to describe this transient phonon hydrodynamic phenomenon, which is named as second sound (or heat waves)~\cite{gurevichShklov1967seonds,Nielsen1970HEATTA,huberman_observation_2019,PhysRevLett_secondNaF,PhysRevLett.28.1461,RevModPhysJoseph89,cattaneo1948sulla,tzou1995a}, many macroscopic heat conduction equations are derived based on kinetic theory~\cite{sussmann1963,WangMr15application,PhysRevB_SECOND_SOUND,PhysRevB.10.3546,PhysRevX.10.011019,de_tomas_kinetic_2014}, such as the Cattaneo-Vernotte equation~\cite{cattaneo1948sulla}, phase lag model~\cite{tzou1995a,tzou1995,tzou1995b}, G-K equation~\cite{PhysRev_GK,PhysRev.148.766}.
Different from Fourier's law with infinite thermal perturbation speed, these hyperbolic heat equations consider the thermal wave propagation with finite speed as well as temperature damping~\cite{PhysRevLett.125.265901,RevModPhysJoseph89,cattaneo1948sulla}.
This novel phonon hydrodynamic phenomenon was observed in cryogenic temperature in bulk materials several decades ago, such as solid helium~\cite{PhysRevLett.16.789}, NaF~\cite{PhysRevLett_secondNaF,PhysRevLett_ssNaf}, Bi~\cite{PhysRevLett.28.1461} and SrTiO$_3$~\cite{PhysRevLett.99.265502}.
Recently, based on the first-principle calculation and linearized phonon BTE, some researchers predicted that the N scattering keeps dominating over R scattering and the second sound can be predicted well-above cryogenic conditions in graphene~\cite{lee_hydrodynamic_2015,cepellotti_phonon_2015}, which has a wider temperature window of hydrodynamic phonon transport compared to those in bulk materials.
Based on phonon BTE under Callaway model, it was found that the speed of second sound is around $4000$ m/s in a $(20,20)$ single-wall carbon nanotube (SWCNT) and the second sound can propagate more than $10~\mu$m in an isotopically pure $(20,20)$ SWCNT for frequency around $1$ GHz at $100$ K~\cite{lee2017}.
Furthermore, the second sound was also directly observed in graphite at temperatures above $100$ K by transient thermal grating (TTG) experiments at micron scale~\cite{huberman_observation_2019}.

\begin{figure}[htb]
 \centering
 \includegraphics[scale=0.38,clip=true]{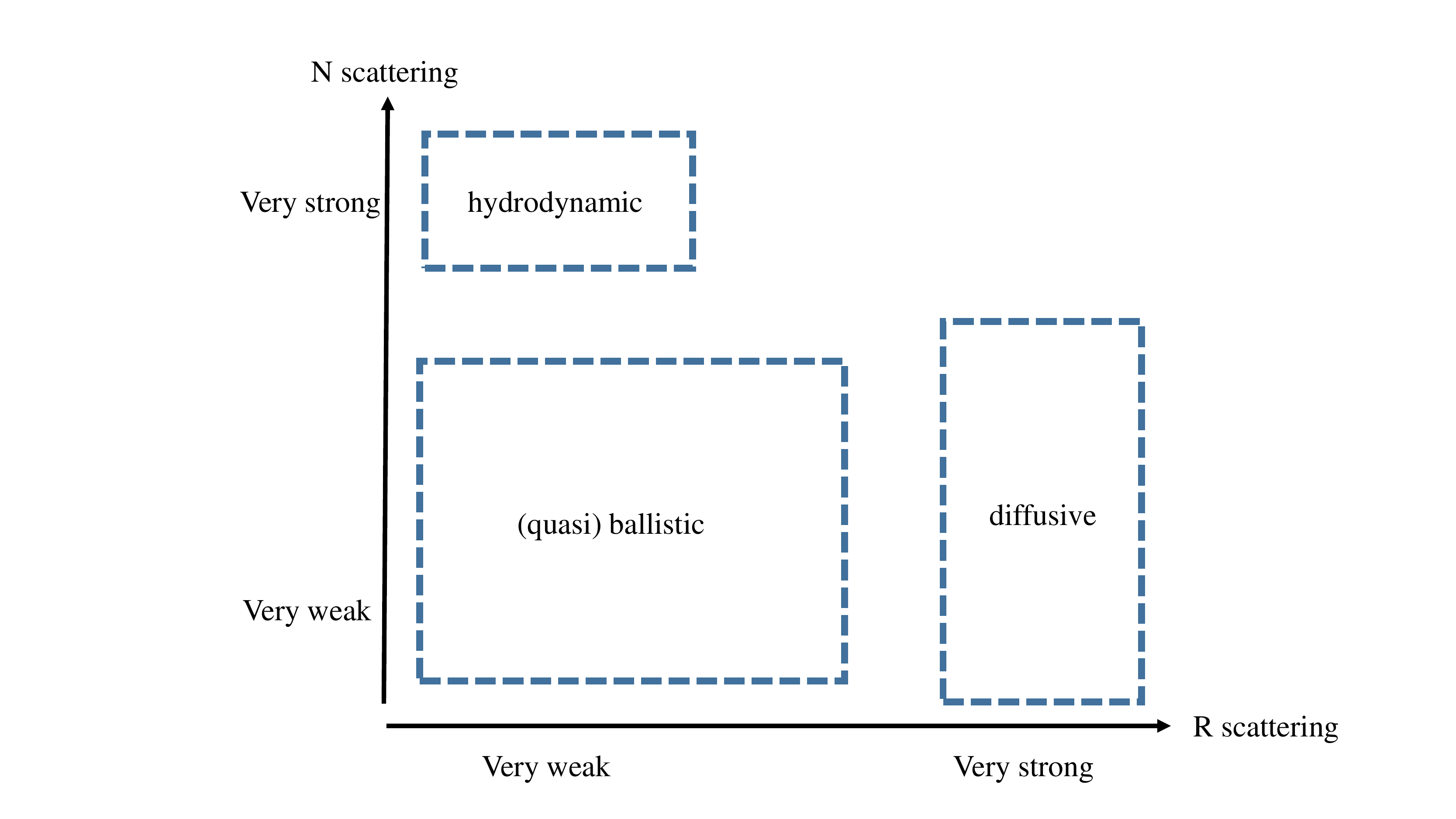}
 \caption{A simple schematic of the main phonon transport regimes mentioned in this work. Diffusive~\cite{ChenG05Oxford}: R scattering happens frequently and dominates heat conduction. (Quasi) ballistic~\cite{PhysRevB.102.104310,PhysRevB.89.094302}: not sufficient phonon (N/R) scattering. Hydrodynamic~\cite{PhysRev_GK,PhysRev.148.766,huberman_observation_2019,PhysRevLett_secondNaF,PhysRevLett.28.1461}: strong N scattering and weak R scattering. }
 \label{quasiballistic}
\end{figure}

Apart from hydrodynamic phonon transport~\cite{huberman_observation_2019,PhysRevLett_secondNaF,PhysRevLett.28.1461,RevModPhysJoseph89}, some studies also predicted the wave like propagation of heat when the characteristic length/time of the thermal system is comparable to or smaller than the phonon mean free path/relaxation time~\cite{kovacs2018,second_sound_ge2020,PhysRevB.101.075303,PhysRevB.72.125413}.
In this regime, the (quasi) ballistic phonon transport~\cite{PhysRevB.102.104310,PhysRevB.89.094302} (\cref{quasiballistic}) dominates heat conduction and the phonon scattering is not sufficient, which indicates that the momentum is almost conserved, too.
For example, in typical TTG studies of suspended silicon membrane~\cite{collins_non-diffusive_2013}, for short TTG periods compared to the phonon mean free path, the model yields a non-exponential or wave like decay of grating amplitude with time rather than an exponential decay in the diffusive regime.
In frequency-domain thermoreflectance (FDTR) experiments~\cite{PhysRevB.101.075303}, when the frequency of time varying heat source increases, the experimental phase shifts and the amplitude of the temperature oscillations can also be observed, which can not be explained by the modified Fourier's law with an effective thermal conductivity.
Recently, the wave like propagation of heat has also been observed in a rapidly varying temperature field in Germanium between $7$ K and room temperature, where the normal scattering is not strong~\cite{second_sound_ge2020}.

It seems that the wake-like propagation of heat can be observed in both the hydrodynamic~\cite{huberman_observation_2019,PhysRevLett_secondNaF,PhysRevLett.28.1461} and (quasi) ballistic regimes~\cite{kovacs2018,second_sound_ge2020,PhysRevB.101.075303}.
Recently, the previous NaF experiments~\cite{PhysRevLett_secondNaF,PhysRevB_NaF71} of second sound and ballistic phonon propagation were quantitatively analyzed~\cite{kovacs2018}.
The results show that the available experimental data is not the best and some conditions indicate possible crucial problems.
According to previous studies and disputes~\cite{RevModPhysJoseph89,PhysRev_GK,PhysRev.148.766,PhysRevB.101.075303,second_sound_ge2020}, it is hard to clearly distinguish the (quasi) ballistic or hydrodynamic phonon transport only by the wave like propagation of heat.
So can we provide an approach to distinguish them?
Or can we find a unique transient heat propagation phenomenon in the phonon hydrodynamic regime?

In this work, the transient heat propagation in homogeneous system is studied based on the phonon Boltzmann transport equation (BTE) under the Callaway model~\cite{luo2019,wangmr17callaway,lee_hydrodynamic_2015,Nanalytical}.
Given a quasi-one or quasi-two (three) dimensional thermal transport problem with homogeneous environment temperature, at initial moment, a heat source is added suddenly at the center of the system with high temperature, then the heat propagates from the center to the outer.
Our results show that in quasi-two (three) dimensional simulations, the transient temperature near the center will be lower than the lowest value of initial environment temperature in the hydrodynamic regime within a certain range of time and space.
This is a unique transient heat conduction phenomenon in the hydrodynamic regime, which will not be observed in quasi-one dimensional simulations.
It cannot be observed if the N scattering does not dominates heat conduction.
Similar phenomenon is also predicted in thermal system with a time varying heat source.

The rest of the paper is organized as follows. In Sec.~\ref{sec:bte}, the basic theory of the phonon BTE is introduced as well as numerical solutions. And then the transient heat propagation with a heat source is studied by the phonon BTE in Sec.~\ref{sec:hotspot} and Sec.~\ref{sec:timevarying}, respectively.
Finally, a conclusion is made in Sec.~\ref{sec:conclusion}.

\textbf{Note:} There are some disputes about the terminology ``hydrodynamic phonon transport" or ``phonon hydrodynamic" in different references~\cite{leesangyeopch1,PhysRevB.101.075303,PhysRevApplied.11.034003,PhysRev_GK,PhysRev.148.766}. {\color{black}{Many researchers have different understandings about this terminology.}}
One of them~\cite{leesangyeopch1,PhysRevB.99.085202,PhysRevLett_secondNaF,nanoletterchengang_2018} focus on the hydrodynamic phenomena of phonon transport due to strong N-scattering, while some researchers~\cite{PhysRevApplied.11.034003,PhysRevApplied.11.054059,PhysRevB.101.075303,PhysRevB.103.L140301} use the hydrodynamic equations to describe the quasiballistic or non-diffusive phonon transport, where the strong N scattering is not required.
Admittedly, this is still an open academic topic.
In this study, the hydrodynamic phonon transport happens when the N scattering is much stronger than the boundary scattering and the boundary scattering is much stronger than the R scattering~\cite{lee_hydrodynamic_2015,cepellotti_phonon_2015,PhysRev_GK,PhysRev.148.766}.

\section{Phonon BTE and numerical solution}
\label{sec:bte}

\subsection{Phonon BTE}

In order to well capture the transient heat propagation in different regimes, the phonon Boltzmann transport equation (BTE) is used and the complex phonon scattering term is simplified by the Callaway's dual relaxation time model~\cite{PhysRev_callaway,wangmr17callaway,PhysRevB.99.085202,luo2019,lee_hydrodynamic_2015}.
{\color{black}{Note that most of previous experiments of hydrodynamic phonon transport are conducted on three-dimensional bulk materials~\cite{PhysRevLett.99.265502,PhysRevLett.28.1461,PhysRevLett_secondNaF,PhysRevLett_ssNaf,huberman_observation_2019}. The experimental proof of transient hydrodynamic phonon transport in low-dimensional materials (e.g., graphene) is still very challenging and difficult.
In this study, we only consider the three-dimensional materials and the wave vector space (or solid angle space) in the phonon BTE simulations are always three-dimensional.
The phonon dispersion is not accounted and the Debye model is used, which is accurate according to previous studies of hydrodynamic phonon transport in bulk materials~\cite{kovacs2015,Nanalytical,nie2020thermal,kovacs2018,Dreyer1993,WangMr15application,PhysRev_GK,PhysRev.148.766}, for example, NaF.}}
The temperature difference in the domain is assumed to be very small compared to the reference temperature $T_0$, i.e., $|T-T_0| \ll T_0$.

Under these simplifications, the phonon BTE can be written as~\cite{Nanalytical,nie2020thermal,shang_heat_2020,wangmr17callaway,luo2019,WangMr15application}
\begin{align}
\frac{\partial e}{\partial t }+ v_g \bm{s} \cdot \nabla_{\bm{x}} e  &= \frac{e^{eq}_{R} -e}{\tau_{R}} + \frac{e^{eq}_{N} -e}{\tau_{N}}, \label{eq:BTE}  \\
e^{eq}_{R} (T) &= \frac{C (T-T_0) }{ 4 \pi},  \label{eq:eeqr}  \\
e^{eq}_{N} (T) &= \frac{C (T-T_0) }{ 4 \pi} +\frac{ C T}{4 \pi} \frac{\bm{s} \cdot \bm{u} }{v_g}, \label{eq:eeqn}
\end{align}
where $e=e(\bm{x},\bm{s},t)$ is the deviational phonon distribution function of energy density, which depends on spatial position $\bm{x}$, unit directional vector $\bm{s}$ and time $t$.
$e^{eq}_R$ and $e^{eq}_N$ are the equilibrium distribution function of the R scattering and N scattering linearized by the specific heat $C=C(T_0)$, respectively.
$\tau_R$ and $\tau_N$ are the relaxation time of the R scattering and N scattering, respectively.
$v_g$ is the value of group velocity, $\bm{u}$ is the drift velocity~\cite{Nanalytical,nie2020thermal,shang_heat_2020}.
In this study, $v_g$, $C$, $\tau_R$ and $\tau_N$ are constants.

Although the phonon dispersion and polarization are not accounted, the model equation~\eqref{eq:BTE} can still provide insight on multiscale thermal transport behaviors, including quasiballistic/hydrodynamic phonon transport or non-equilibrium effects~\cite{cao2007thermomass,Nanalytical,nie2020thermal,shang_heat_2020,MurthyJY05Review}.
Furthermore, the simple expression of phonon BTE helps us to find the essential physical mechanisms of wave like propagation of heat more clearly.

To close the phonon BTE, the conservation principle of the phonon scattering is used.
Both the N scattering and R scattering satisfy the energy conservation so that
\begin{align}
\int  \frac{e^{eq}_{R} -e}{\tau_{R}} d\Omega = \int  \frac{e^{eq}_{N} -e}{\tau_{N}} d\Omega =0,
\label{eq:energyconservation}
\end{align}
where $d\Omega$ is the integral over the whole solid angle space.
Besides, the N scattering conserves momentum~\cite{Nanalytical,nie2020thermal,PhysRev_callaway}, i.e.,
\begin{align}
\int  \bm{s}   \frac{e^{eq}_{N} -e}{\tau_{N}} d\Omega =0.
\label{eq:momentumconservation}
\end{align}

Based on kinetic theory~\cite{ChenG05Oxford}, the macroscopic spatio-temporal distributions including local energy $E(\bm{x},t)$, local temperature $T(\bm{x},t)$ and heat flux $\bm{q}(\bm{x},t)$ can be obtained by taking the moments of phonon distribution function~\cite{wangmr17callaway,PhysRevB.99.085202,luo2019}, i.e.,
\begin{align}
E &= \int   e  d\Omega , \label{eq:energy} \\
\bm{q} &= \int v_g \bm{s}  e  d\Omega .   \label{eq:heatflux} \\
T &= T_0 + E/C .  \label{eq:T}
\end{align}
Note that the temperature $T$ calculated by Eq.~\eqref{eq:T} is not the thermodynamic temperature, but more like the symbol of local energy density~\cite{ChenG05Oxford}.
For the sake of convenience, it is referred to as the temperature in the following discussions.
Based on Eqs.~\eqref{eq:eeqn},~\eqref{eq:momentumconservation} and~\eqref{eq:heatflux}, the drift velocity can be obtained~\cite{WangMr15application,Nanalytical},
\begin{align}
\bm{u} = \frac{3 \bm{q} }{C T }. \label{eq:driftvelocity}
\end{align}

{\color{black}{In addition, in the phonon BTE simulations of three-dimensional materials or thermal systems, when the system length in one direction is infinite (or the heat conduction in some direction can be ignored/reduced), the three-dimensional simulations can be reduced into quasi-one or quasi-two dimensional simulations, such as quasi-one dimensional heat pulse~\cite{kovacs2018,Dreyer1993}/TTG simulations~\cite{collins_non-diffusive_2013} and quasi-two dimensional in-plane heat conduction~\cite{LUO2017970}.
Therefore, in the present work the terminology ``quasi-one'' or ``quasi-two'' does not represent one-dimensional or two-dimensional materials.
}}

\subsection{Dimensional treatments}

We choose $e_{\text{ref}}$, $T_0$, $L_{\text{ref}}$, $v_{g}$, $t_{\text{ref}}$ as the reference variables to normalize the phonon BTE so that Eq.~\eqref{eq:BTE} becomes
\begin{align}
\frac{\partial e^*}{\partial t^* }+ \bm{s} \cdot \nabla_{\bm{x}^*} e^*  &= \frac{e^{eq,*}_{R} -e^*}{\text{Kn}_{R}} + \frac{e^{eq,*}_{N} -e^*}{\text{Kn}_{N}}, \label{eq:dimensionlessBTE}
\end{align}
where
\begin{align}
e^* &= \frac{e}{e_{\text{ref}} }, &\quad  e^{eq,*}_R &= \frac{e^{eq}_R}{e_{\text{ref}} }, &\quad  e^{eq,*}_N &= \frac{e^{eq}_N}{e_{\text{ref}} } , \notag  \\
e_{\text{ref}}  &=\frac{ C  \Delta T  }{4 \pi}, &\quad  \bm{x^*} &=\frac{\bm{x} }{L_{\text{ref}} }, &\quad  t^* &=  \frac{ t }{t_{\text{ref}} },   \notag  \\
t_{\text{ref}}& = L_{\text{ref}} /v_g, &\quad \text{Kn}_{R} &= \frac{v_g \tau_R}{ L_{\text{ref}} }, &\quad \text{Kn}_{N} &= \frac{v_g \tau_N}{ L_{\text{ref}} }, \label{eq:dimensionlessparameters}
\end{align}
where $\Delta T$ is the temperature difference in the domain.
In the absence of time-varying heat source or oscillatory boundary, based on dimensional analysis~\cite{barenblatt1987dimensional}, it can be found that the transient heat propagation predicted by phonon BTE is mainly determined by two Knudsen numbers ($\text{Kn}_{R}$ and $\text{Kn}_{N}$), which represent the ratio between the phonon mean free path of the R scattering and N scattering to the characteristic length.
Note that if a time varying heat source appears and the its heating-period is much smaller than $t_{\text{ref}}$~\cite{PhysRevB.101.075303,second_sound_ge2020}, the characteristic time should be the heating-period of the heat source, which is not considered in the present work.

\subsection{Numerical solution}

The discrete unified gas kinetic scheme (DUGKS)~\cite{GuoZl13DUGKS}, which is a finite volume method and has been successfully applied in multiscale phonon transport~\cite{guo_progress_DUGKS,GuoZl16DUGKS,LUO2017970,zhang_discrete_2019,luo2019}, is used to solve the dimensionless phonon BTE~\eqref{eq:dimensionlessBTE} in the whole discretized phase space.
The physical space is discretized with $N_{cell}$ uniform cells, and the time step of the DUGKS is
\begin{align}
\Delta t= \text{CFL} \times \frac{\Delta x}{v_g},
\end{align}
where $\Delta x$ is the discretized cell size and $\text{CFL}$ is the Courant–Friedrichs–Lewy number.
The first-order upwind scheme is used to deal with the spatial gradient of the distribution function in the ballistic regime, and the van Leer limiter~\cite{vanleer1977} is used in the hydrodynamic and diffusive regimes.
For the discretization of the solid angle space $\bm{s}= \left( \cos \theta, \sin \theta \cos \varphi, \sin \theta \sin \varphi \right)$, where $\theta$ is the polar angle and $\varphi$ is the azimuthal angle, $\cos \theta$ is discretized with the $N_{\theta}$-point Gauss-Legendre quadrature~\cite{NicholasH13GaussL}, while the azimuthal angle $\varphi \in [0,\pi]$ (due to symmetry) is discretized with the $N_{\varphi}$-point Gauss-Legendre quadrature.
The thermalizing boundary condition~\cite{GuoZl16DUGKS} is used to deal with the system boundary $\bm{x}_b$.
It assumes that all phonons coming from boundaries are the equilibrium state with boundary temperature $T(\bm{x}_b)$, i.e.,
\begin{align}
e(\bm{x}_b,\bm{s})=e^{eq}_R (T(\bm{x}_b)). \label{eq:BC}
\end{align}
More detailed introductions of the numerical solutions can be found in Ref~\cite{luo2019}.

\section{Heat propagation with a hotspot}
\label{sec:hotspot}

\subsection{Problem description}

\begin{figure}[htb]
 \centering
 \subfloat[] { \includegraphics[scale=0.5,viewport=50 50 450 390,clip=true]{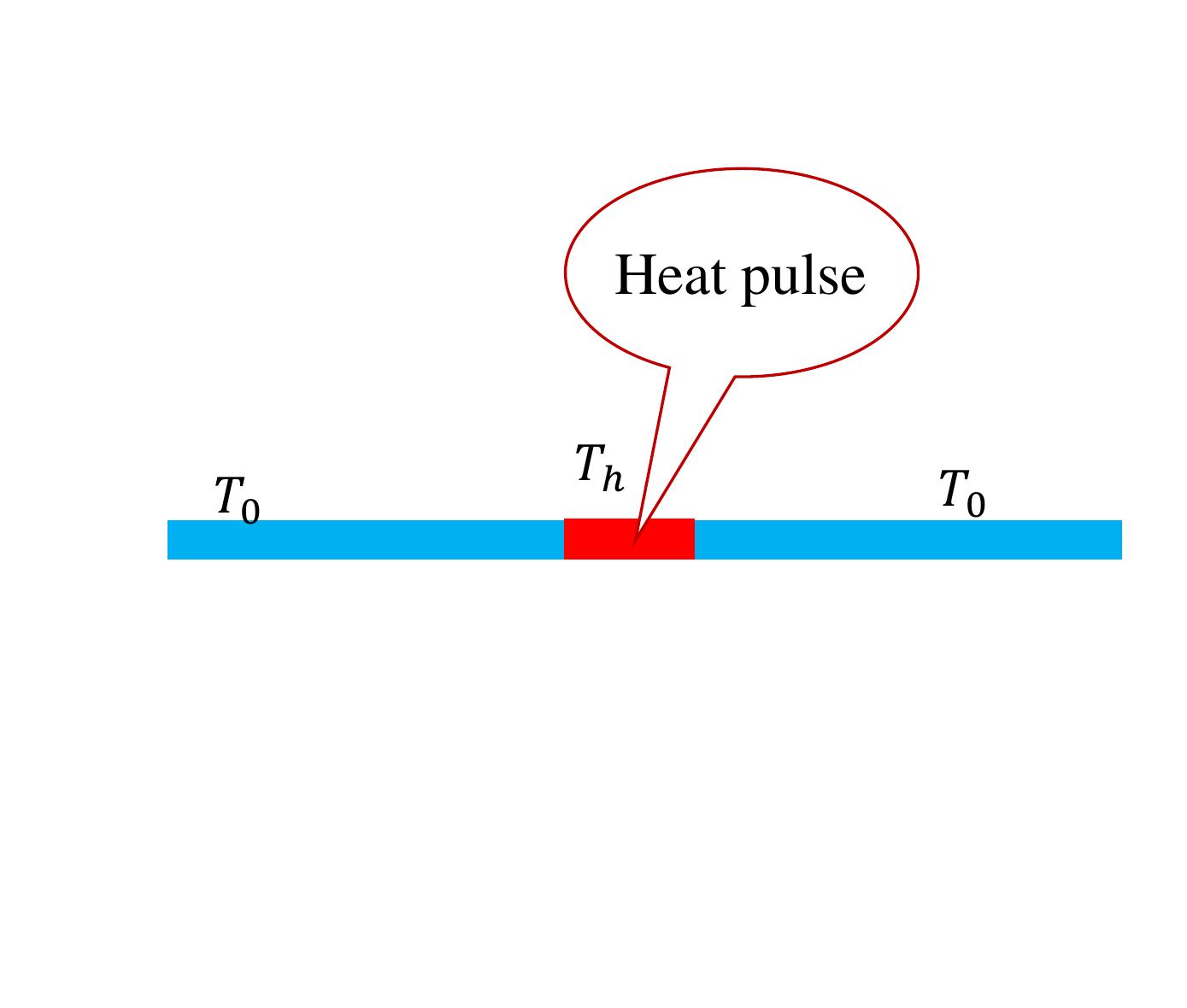} }~~
 \subfloat[] { \includegraphics[scale=0.5,viewport=500 50 800 390,clip=true]{problem_12D.pdf} }
 \caption{In homogeneous system with environment temperature $T_0$, at initial moment $t=0$, a heat pulse is added suddenly and a hotspot with high temperature $T_h$ appears at the center of the domain. As $t>0$, the heat pulse disappears and heat propagates from the center to the outer, where $T_h > T_0$.    (a) Quasi-one dimensional simulation. (b) Quasi-one dimensional simulation. {\color{black}{Actually, from the perspective of 3D, the quasi-two dimensional simulation is like a concentric cylindrical structure. While the three-dimensional heat propagation is like a concentric spherical shells structure. The diameters of hotspot and computational domain are $L_h$ and $L=10 L_h$, respectively. }}  }
 \label{problemdescription}
\end{figure}

\begin{figure}[htb]
 \centering
 \includegraphics[scale=0.3,clip=true]{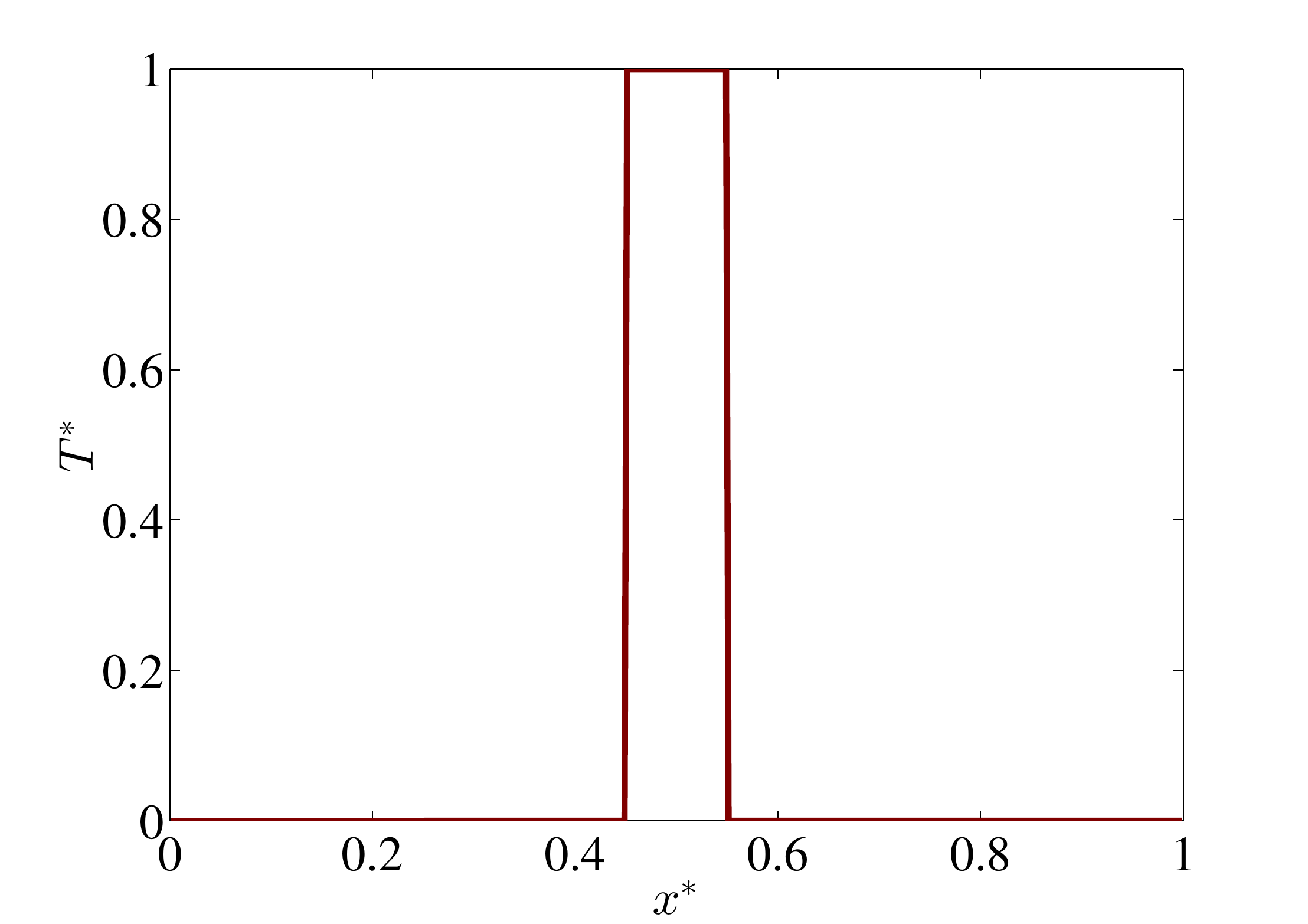}
 \caption{The initial temperature distribution along a line across the center of hotspot, where $t=0$, $T^*=(T-T_0)/(T_h -T_0)$, $x^*=x/L$, $x^*=0.5$ is the center of the computational domain (or hotspot). }
 \label{initialtemperature}
\end{figure}

Given a homogeneous system with environment temperature $T_0$, at initial moment $t=0$, a heat pulse is added on this system suddenly and a hotspot with high temperature $T_h$ appears at the center of the domain.
The schematic of the transient heat propagation problem is shown in~\cref{problemdescription}.
In quasi-one dimensional simulations, the width of the hotspot at $t=0$ is $L_h$ and the total length of the computational domain is $L=10 L_h$.
In quasi-two (or three) dimensional simulations, the diameters of hotspot and computational domain are $L_h$ and $L=10 L_h$, respectively.
In this simulations, we set $\Delta T= T_h -T_0 \rightarrow  0^{+}$, $L_{\text{ref}} =L$.
The temperature distributions in the whole domain at $t=0$ is shown in~\cref{initialtemperature}, where $T^*=(T-T_0)/(T_h -T_0)$ is the normalized temperature and $x^*=x/L$ is the normalized coordinates along a line across the center of hotspot $\bm{x}_0$.
When $t>0$, the heat propagates from the center to the outer.

\subsection{Theoretical analysis in the ballistic and diffusive limits}
\label{ballistic_diffusive}

In the diffusive limit $(\text{Kn}_R \rightarrow 0,\text{Kn}_N = \infty)$, the Fourier's law of thermal conduction is valid so that the temperature decays exponentially with time.
The transient temperature will not be lower than the lowest temperature at the initial moment, i.e., $T \geq T_0$.

In the ballistic limit $(\text{Kn}_R = \infty,\text{Kn}_N = \infty)$, phonon mitigates inside the system without scattering so that Eq.~\eqref{eq:BTE} becomes
\begin{align}
\frac{\partial e}{\partial t }+ v_g \bm{s} \cdot \nabla_{\bm{x}} e &=0,  \notag \\
\Rightarrow  e( \bm{x}, \bm{s}, t+ t_1 ) &= e(  \bm{x}- v_g \bm{s} t_1, \bm{s}, t),   \label{eq:ballistic}
\end{align}
where $t_1$ is an arbitrary time interval.
Based on Eq.~\eqref{eq:ballistic}, it can be found that the distribution function does not vary with spatial position.
Note that the local energy density (or temperature) is an integral of the phonon distribution function over the whole solid angle space, i.e., Eq.~\eqref{eq:energy} (or Eq.~\eqref{eq:T}), so that we can get
\begin{align}
T(\bm{x},t)= T_0 + \frac{1}{C}  \int e(\bm{x},\bm{s},t) d\Omega =T_0 + \frac{1}{C}  \int e(\bm{x}-\bm{s} t,\bm{s}, 0) d\Omega .
\label{eq:Tballistic}
\end{align}
Based on the initial settings, i.e.,
\begin{align}
\begin{split}
e(\bm{x},\bm{s}, 0) &= \left \{
\begin{array}{ll}
   e^{eq}_R (T_h) ,                    &  r  \leq  L_h/2, \\
   e^{eq}_R (T_0) ,     & r > L_h/2 ,
\end{array}
\right.
\end{split}
\label{eq:initial_distribution}
\end{align}
where $r=\left\| \bm{x}- \bm{x}_0 \right\|_2 $ is the distance from the center of the hotspot $\bm{x}_0$, we can obtain $T(\bm{x},t) \geq T_0$ (or $T^*( \bm{x^*},t^*) \geq 0$), namely, the local temperature at any position or any time will not be lower than the lowest temperature at the initial moment in the ballistic limit.

\subsection{Quasi-one dimensional heat propagation}
\label{sec:quasionecase1}

\begin{figure}[htb]
 \centering
 \subfloat[$\text{Kn}_{R}= \infty$,$\text{Kn}_{N}=10.0$]{\includegraphics[scale=0.26,clip=true]{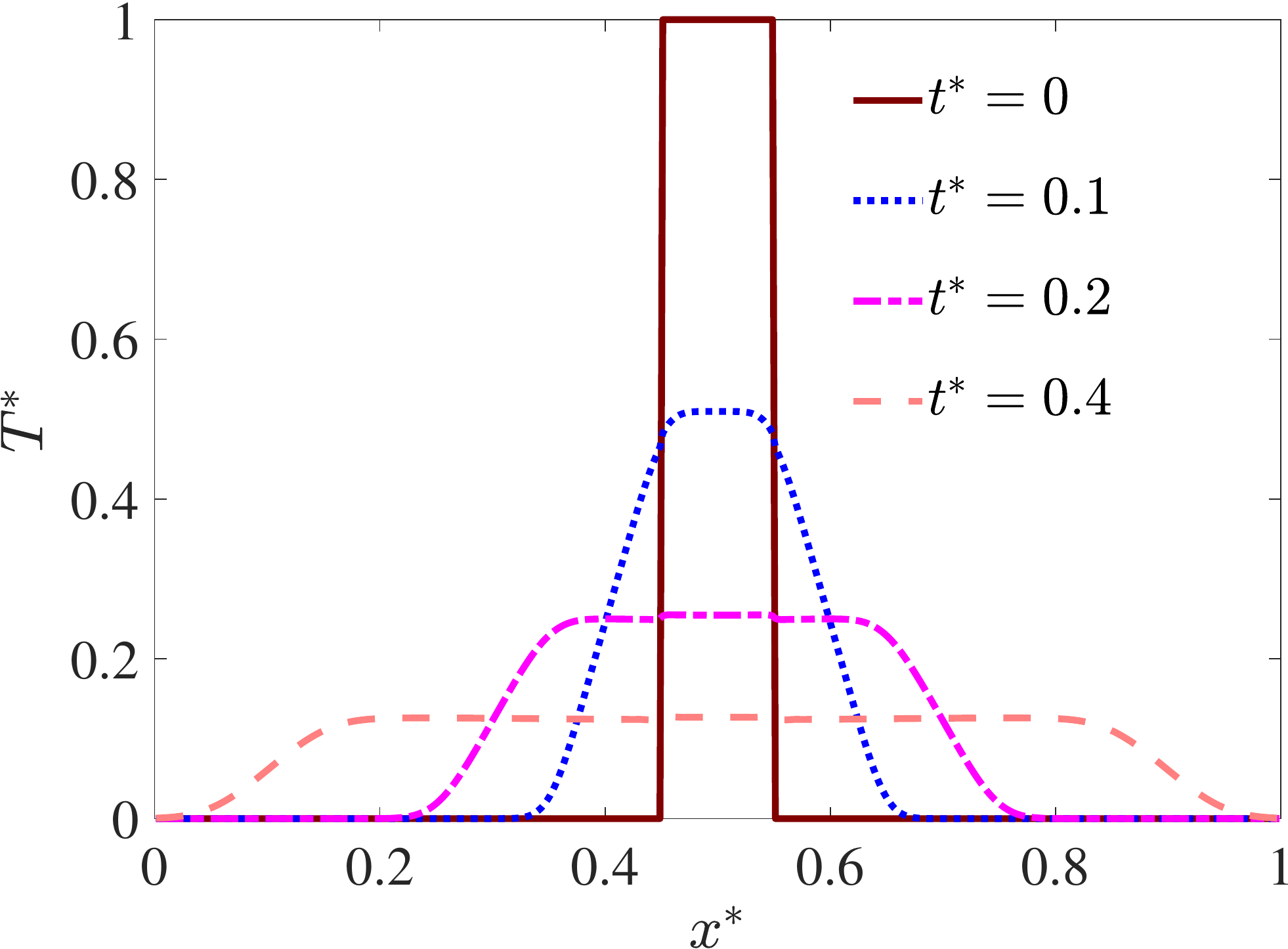}} ~
 \subfloat[$\text{Kn}_{R}= \infty$,$\text{Kn}_{N}=1.0$]{\includegraphics[scale=0.26,clip=true]{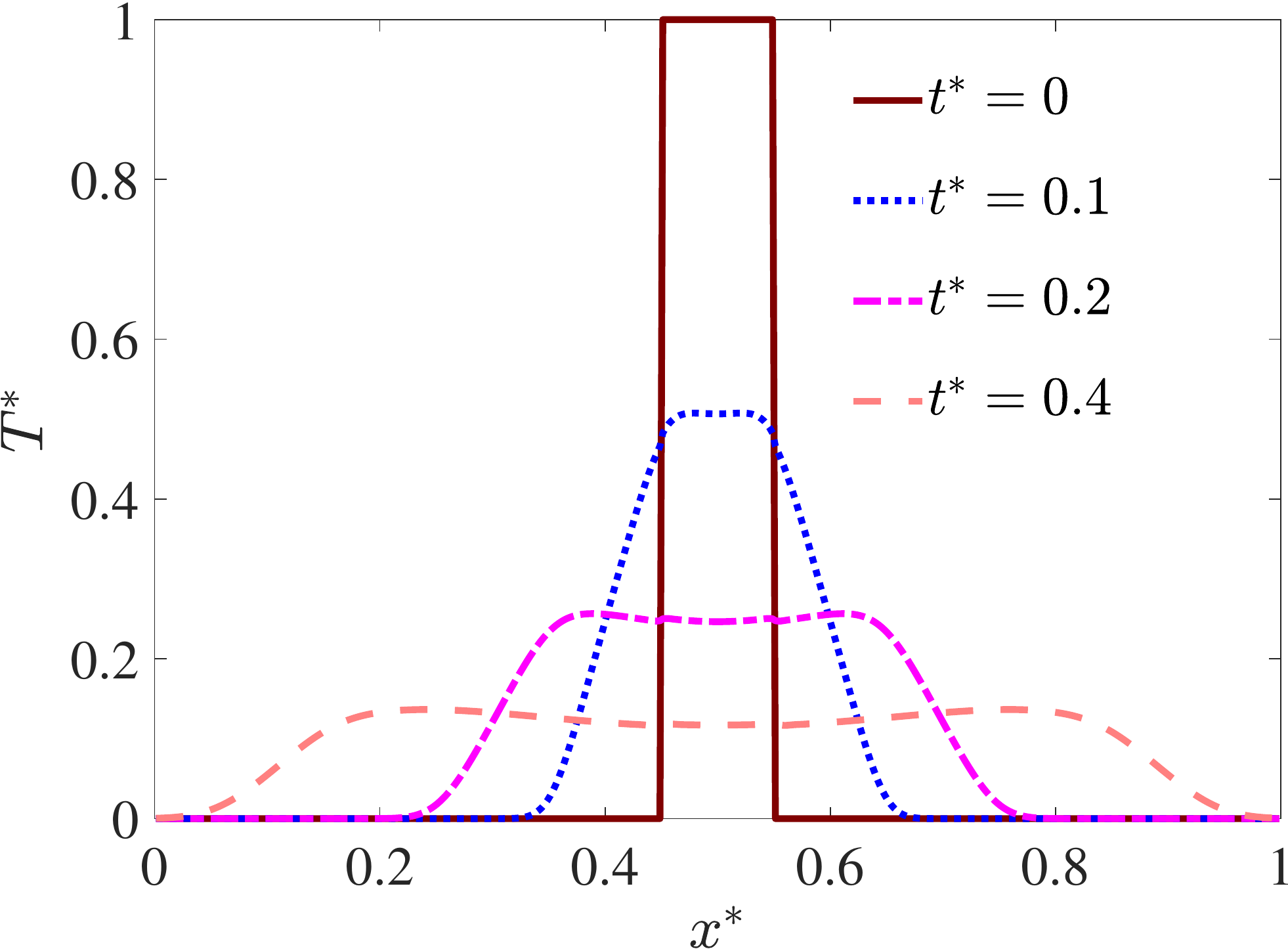}}~
 \subfloat[$\text{Kn}_{R}= \infty$,$\text{Kn}_{N}=0.1$]{\includegraphics[scale=0.26,clip=true]{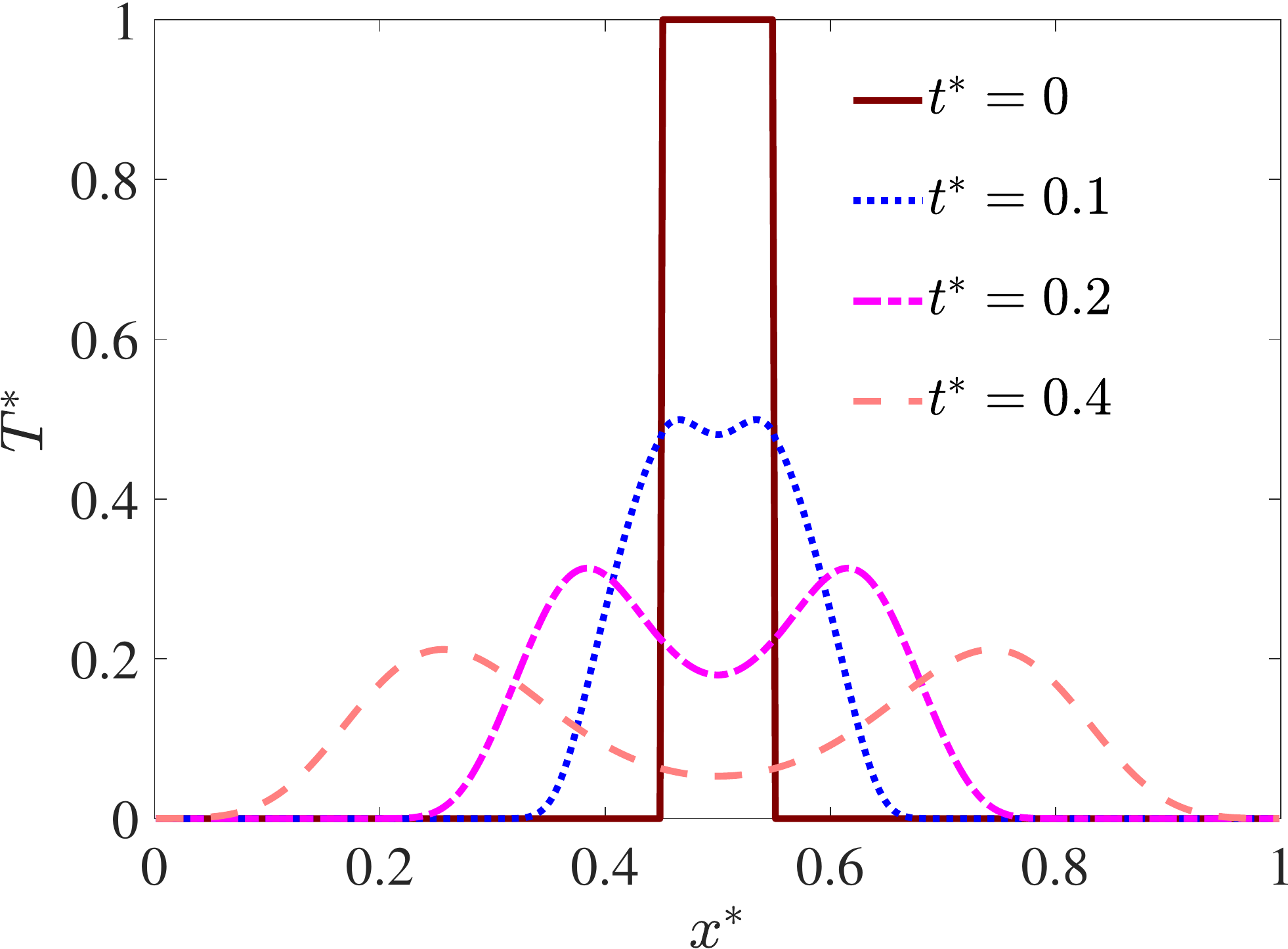}}   \\
 \subfloat[$\text{Kn}_{R}= \infty$,$\text{Kn}_{N}=0.01 $]{\includegraphics[scale=0.26,clip=true]{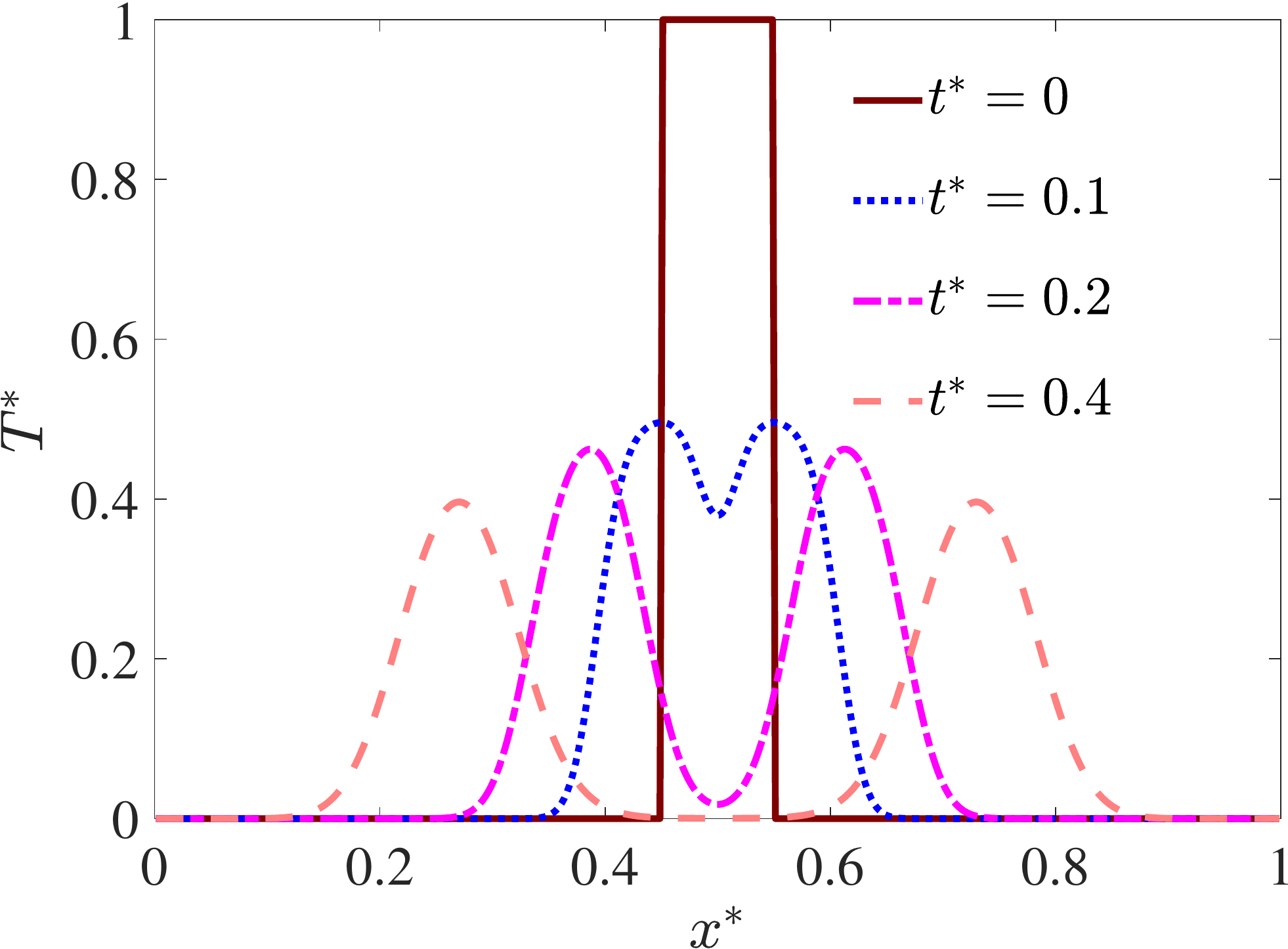}}~
 \subfloat[$\text{Kn}_{R}=0.1$,$\text{Kn}_{N}=0.01$]{\includegraphics[scale=0.26,clip=true]{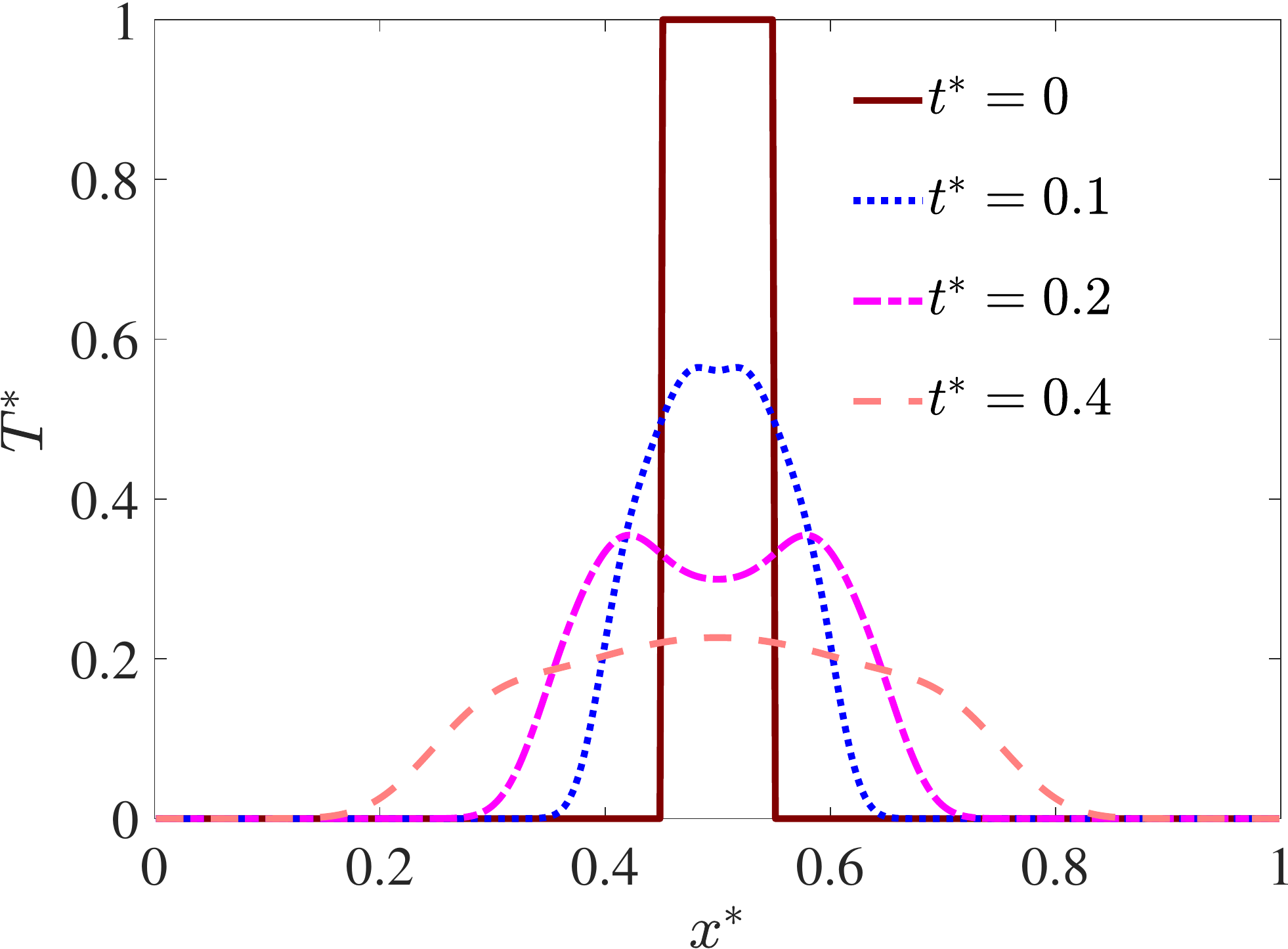}}~
 \subfloat[$\text{Kn}_{R}=0.01$,$\text{Kn}_{N}=0.01$]{\includegraphics[scale=0.26,clip=true]{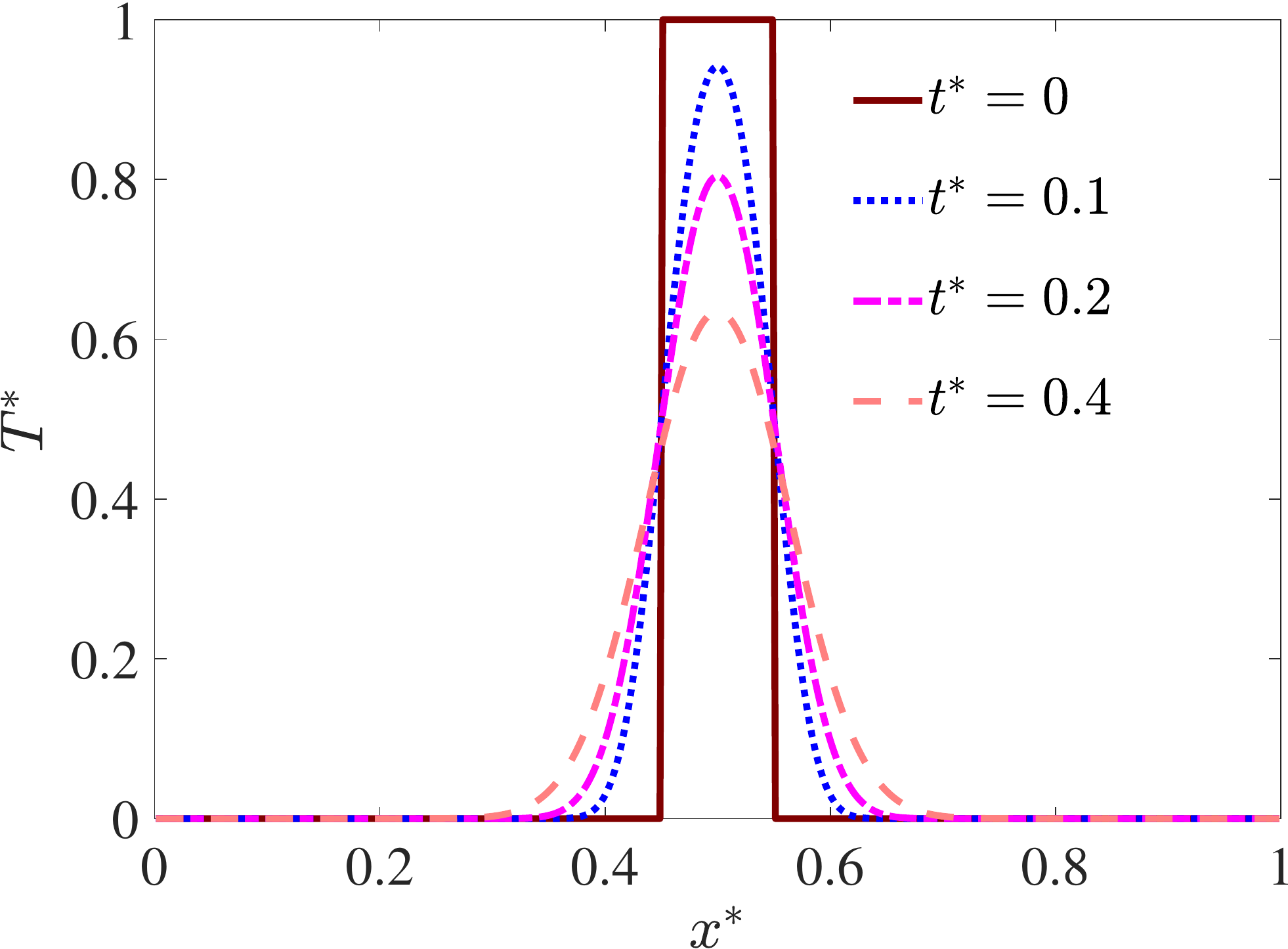}} \\
 \caption{The distributions of spatio-temporal temperature $T^*(x^*,t^*)$ in quasi-one dimensional simulations (\cref{problemdescription}(a)) based on dimensionless phonon BTE (Eqs.~\eqref{eq:dimensionlessBTE},~\eqref{eq:dimensionlessparameters}), where $T^*=(T-T_0)/(T_h -T_0)$, $t^* \in [0,0.45)$, $x^* \in [0,1]$.}
 \label{1DSSUN}
\end{figure}

The quasi-one dimensional wave like propagation of heat (\cref{problemdescription}(a)) is studied with different phonon scattering rates.
We set $N_{cell}=400$, $\text{CFL}=0.2$, $N_{\theta} =100$ and $N_{\varphi}=4$, which is enough to capture the multiscale heat transfer accurately.
In all numerical simulations, $t^* < 0.45$ so that the phonons coming from the hotspot areas will not go out of the computational domain.
The thermalizing boundary condition (Eq.~\eqref{eq:BC}) is used to deal with the boundaries of computational domain.

As shown in~\cref{1DSSUN}(a)(b), in the (quasi) ballistic regime, numerical results (i.e., $T^* \geq 0$) proves that our above theoretical analysis~\ref{ballistic_diffusive} is right.
Besides, it can be found that the heat propagation speed is close to the group velocity $v_g$ (\ref{sec:dugks_validation}).

When the N scattering becomes strong and dominates heat conduction, the phonon transport is in the hydrodynamic regime.
As shown in~\cref{1DSSUN}(c)(d), the wave like propagation of heat appears because the phonon BTE recovers the temperature wave equation in the hydrodynamic limit~\cite{PhysRev_GK,PhysRev.148.766,lee_hydrodynamic_2015,Nanalytical,nie2020thermal,shang_heat_2020} ($\text{Kn}_N \rightarrow 0$, $\text{Kn}_R= \infty$, see \ref{sec:temperature_wave}),
\begin{align}
\frac{\partial^2 T}{\partial t^2 } = \frac{1}{v_{ss} ^2}    \nabla^2_{\bm{x}}  T,
\label{eq:wavetemerpature}
\end{align}
where $v_{ss}=v_g/ \sqrt{3}$ is the speed of second sound.
Note that there is still some decaying of the peak temperature in our simulations.
When $\text{Kn}_{R}= \infty$ and $\text{Kn}_{N}=0.01$, the speed of heat propagation is about $0.52$, which is calculated by the propagation of peak temperature shown in~\cref{1DSSUN}(d).
This value is smaller than $1/ \sqrt{3} \approx 0.58$.
That's because Eq.~\eqref{eq:wavetemerpature} is derived from phonon BTE by assuming $e=e^{eq}_N$, which is impossible in actual thermal systems~\cite{nie2020thermal} (see \ref{sec:temperature_wave} and \ref{sec:dugks_validation}).
As the R scattering becomes strong, from~\cref{1DSSUN}(e)(f), it can be observed that the propagation of heat is impeded and the temperature is significantly damped, because the R scattering does not conserve momentum and causes large heat dissipations.

According to the results in quasi-one dimensional simulations (\cref{1DSSUN}), it can be found that the spatio-temporal temperature will not be lower than the lowest temperature at the initial moment, i.e., $T^*(\bm{x^*},t^*) \geq 0$ or $T(\bm{x},t) \geq T_0$.
In addition, both N scattering and R scattering decreases the speed of heat propagation compared to that in the ballistic regime.
\begin{figure}[htb]
 \centering
 \subfloat[$\text{Kn}_{R}=\infty $,$\text{Kn}_{N}=10.0 $]{\includegraphics[scale=0.26,clip=true]{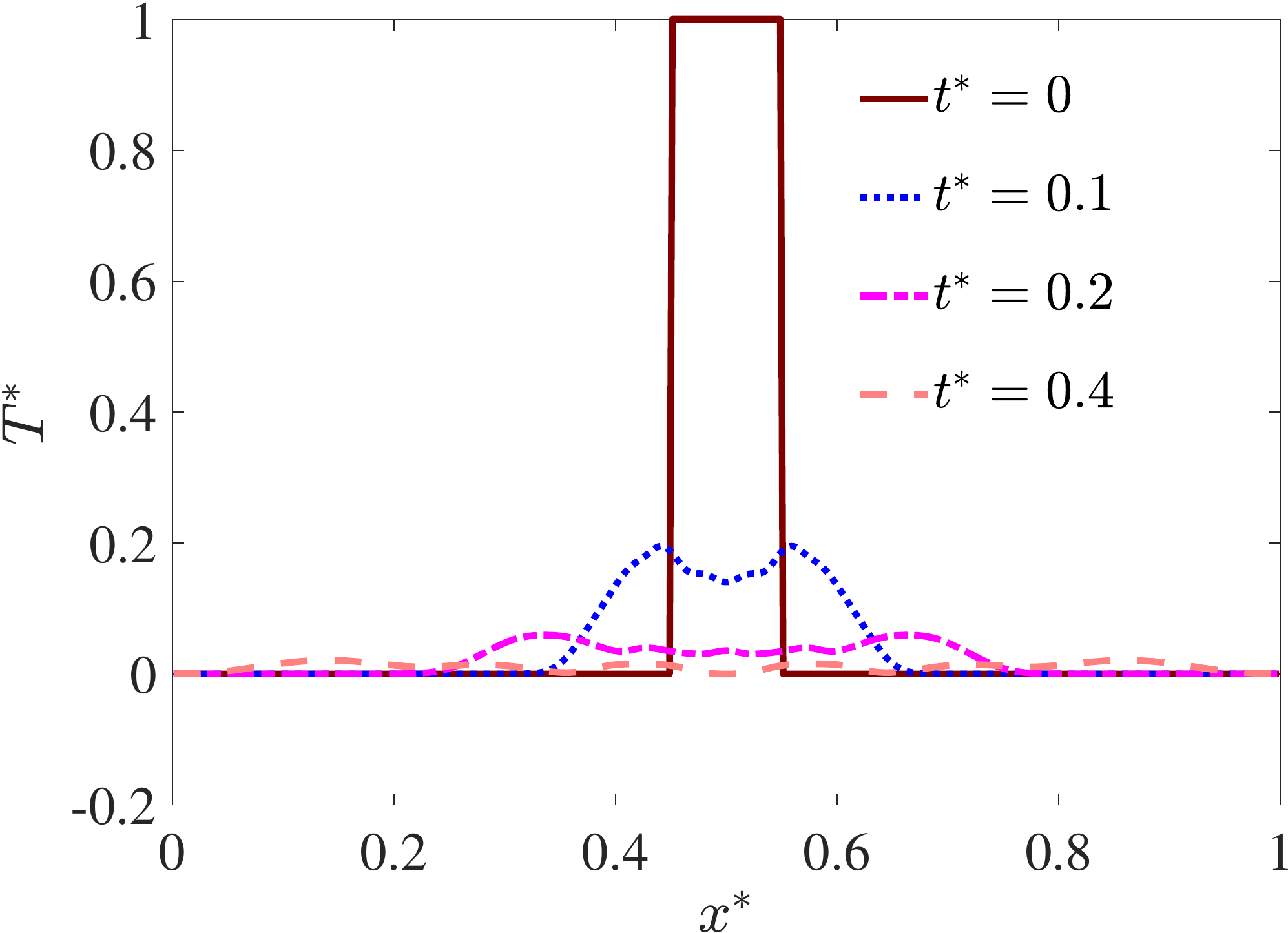}}~
 \subfloat[$\text{Kn}_{R}=\infty $,$\text{Kn}_{N}=1.0 $]{\includegraphics[scale=0.26,clip=true]{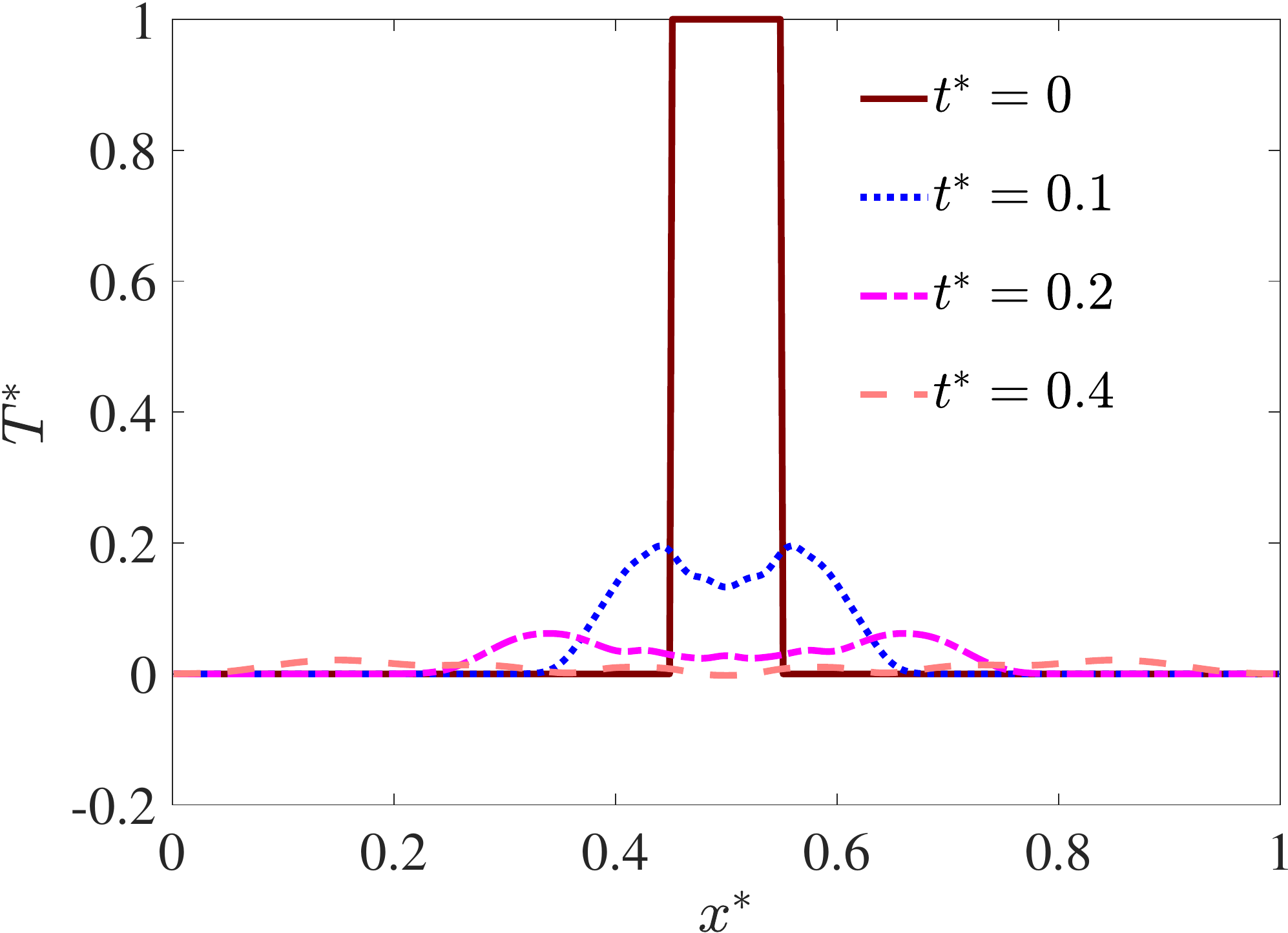}}~
 \subfloat[$\text{Kn}_{R}=\infty $,$\text{Kn}_{N}=0.1 $]{\includegraphics[scale=0.26,clip=true]{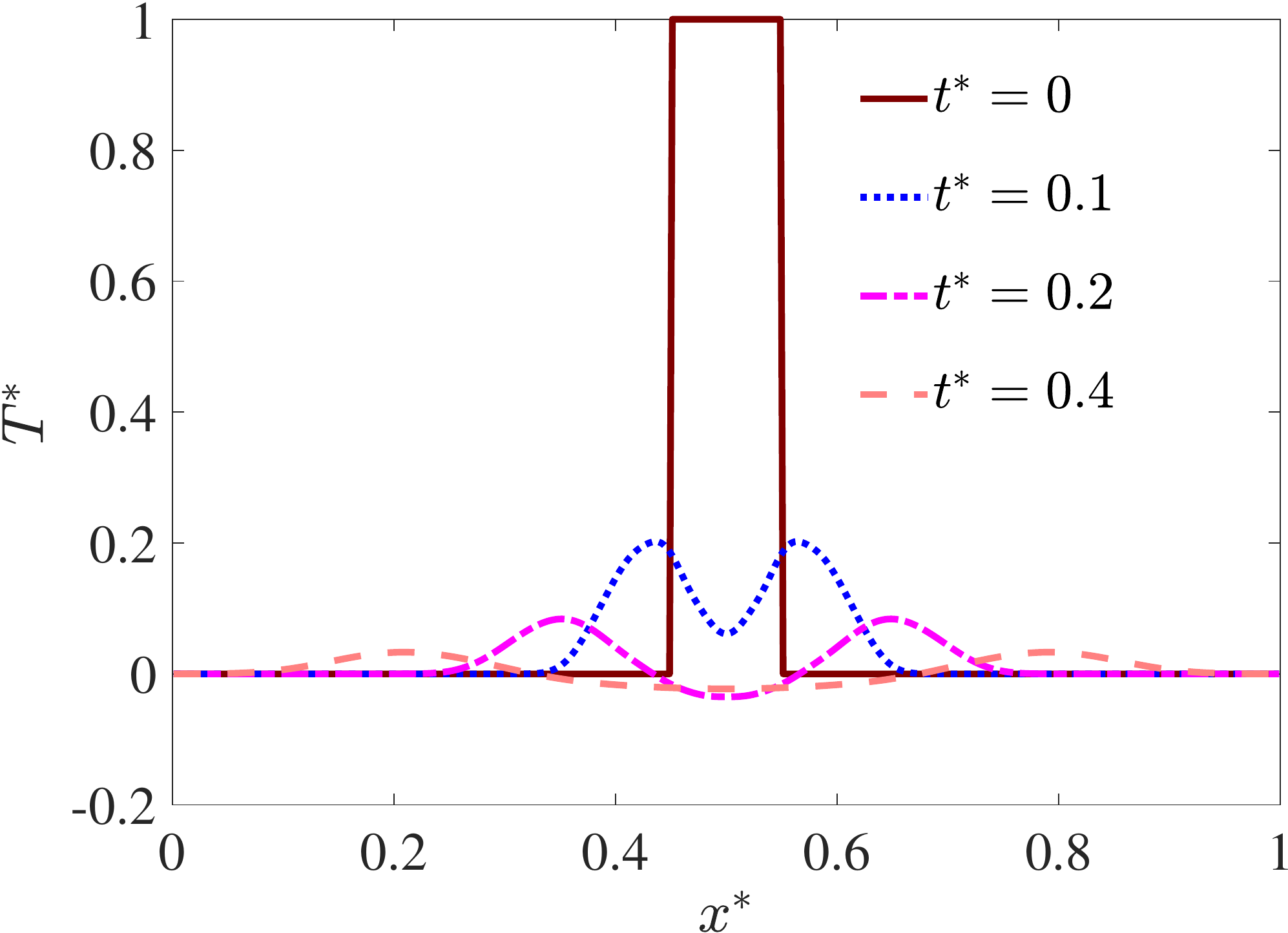}} \\
 \subfloat[$\text{Kn}_{R}=\infty$,$\text{Kn}_{N}=0.01$] {\includegraphics[scale=0.26,clip=true]{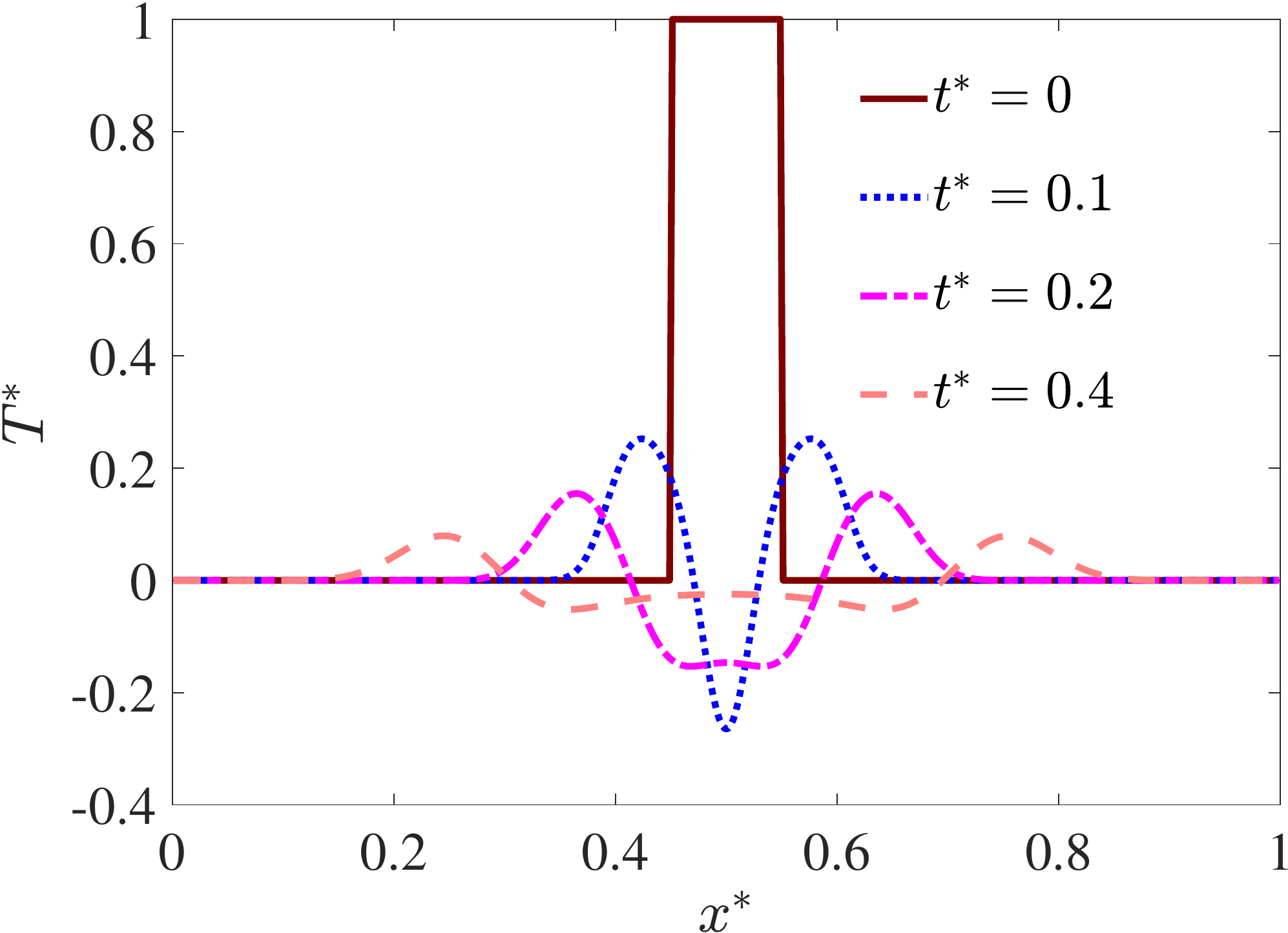}}~
 \subfloat[$\text{Kn}_{R}=1.0 $,$\text{Kn}_{N}=1.0 $] {\includegraphics[scale=0.26,clip=true]{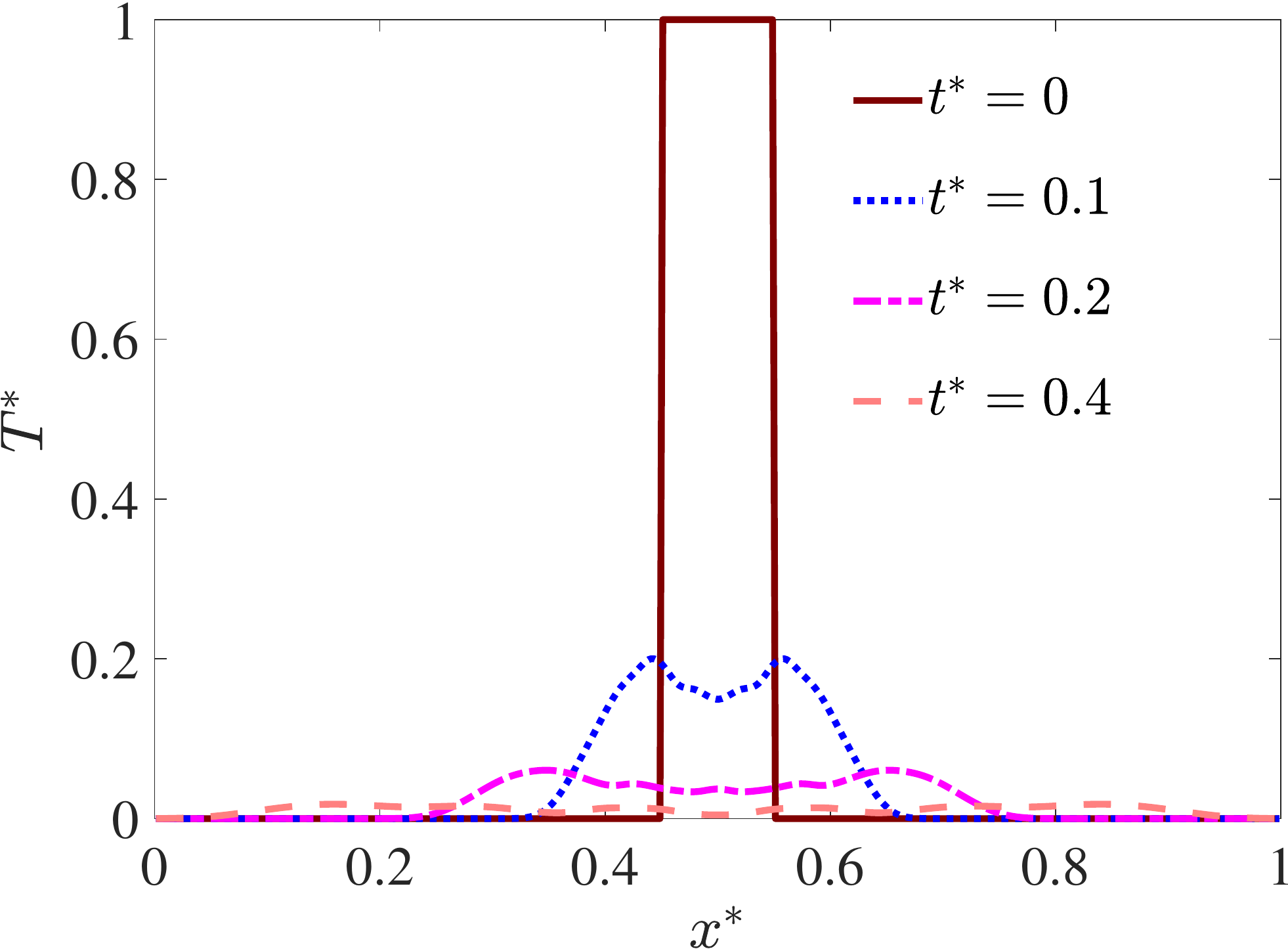}}~
 \subfloat[$\text{Kn}_{R}=1.0 $,$\text{Kn}_{N}=0.1 $]{\includegraphics[scale=0.26,clip=true]{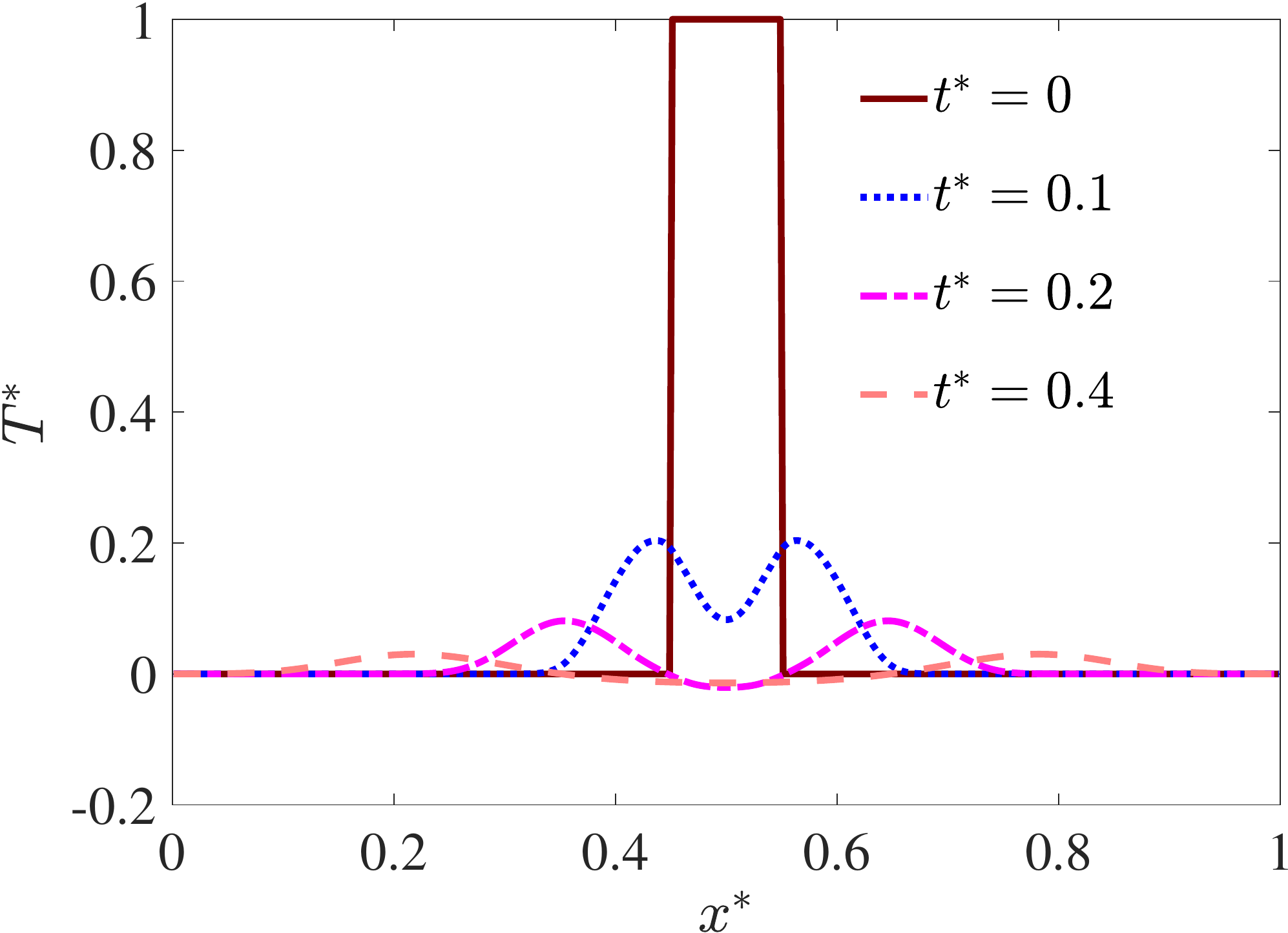}} \\
  \subfloat[$\text{Kn}_{R}= 0.1 $,$\text{Kn}_{N}= 0.1 $]{\includegraphics[scale=0.26,clip=true]{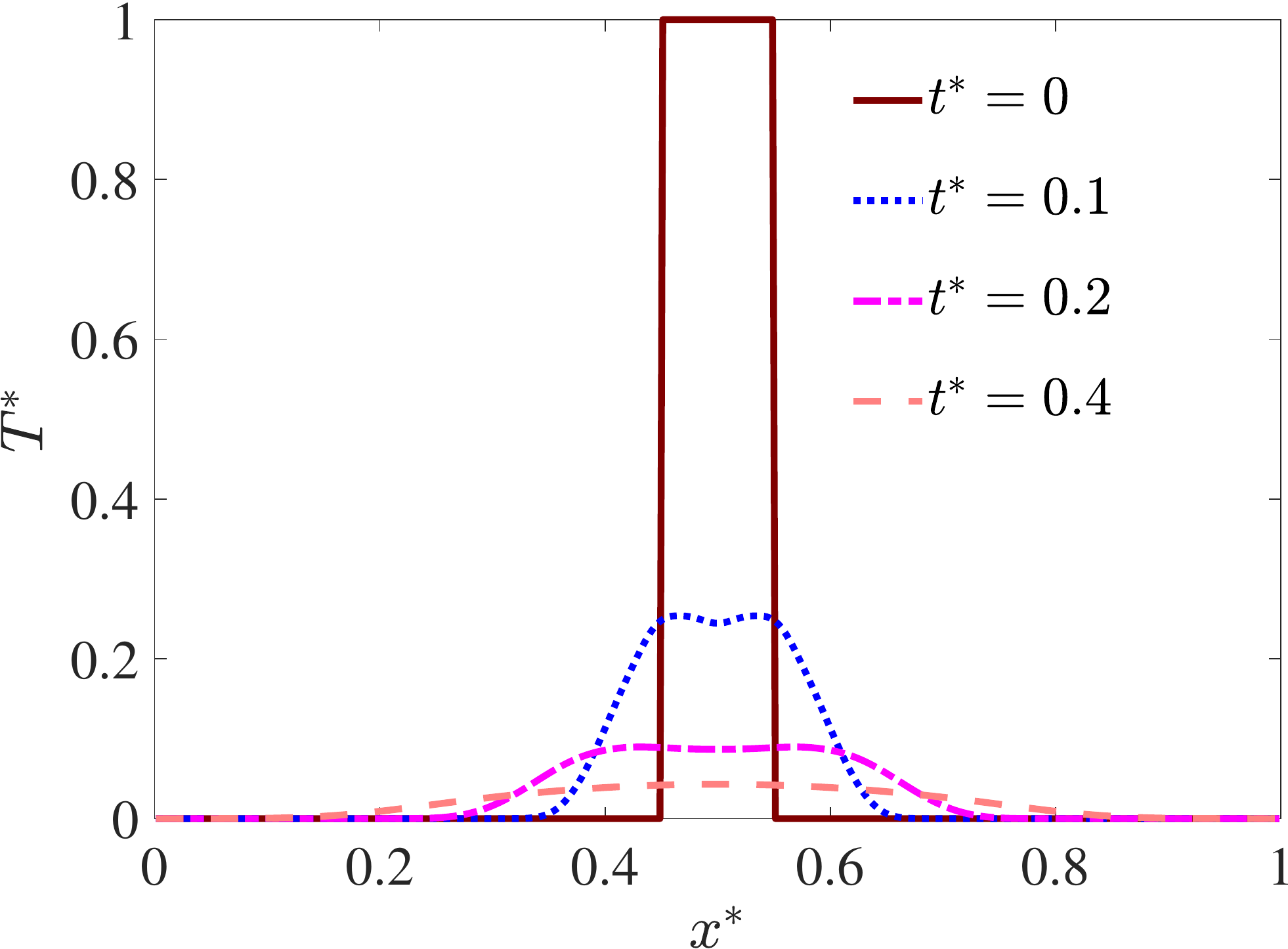}}~
 \subfloat[$\text{Kn}_{R}= 0.1 $,$\text{Kn}_{N}= 0.01 $]{\includegraphics[scale=0.26,clip=true]{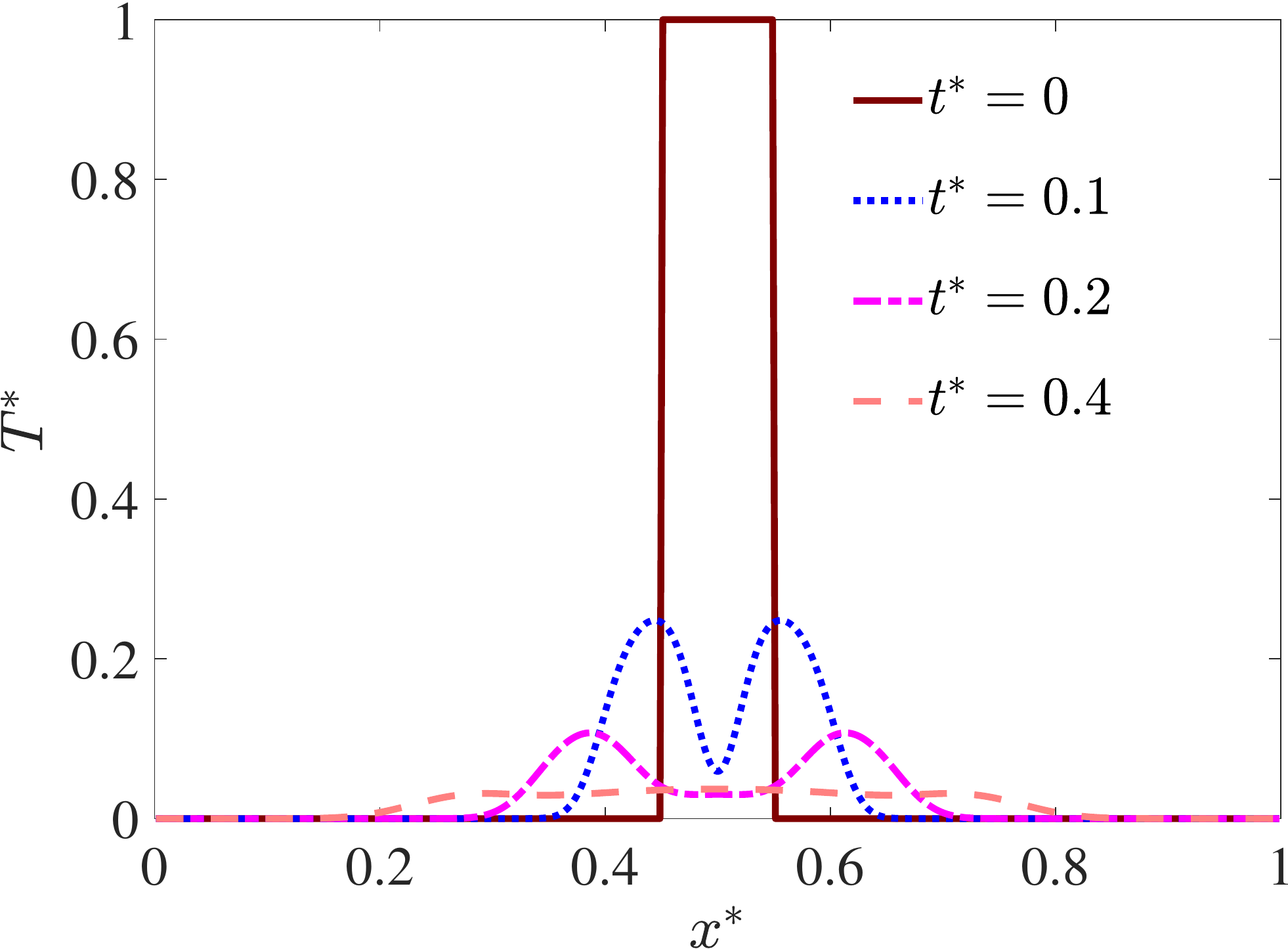}}~
 \subfloat[$\text{Kn}_{R}= 0.01 $,$\text{Kn}_{N}= 0.01 $]{\includegraphics[scale=0.26,clip=true]{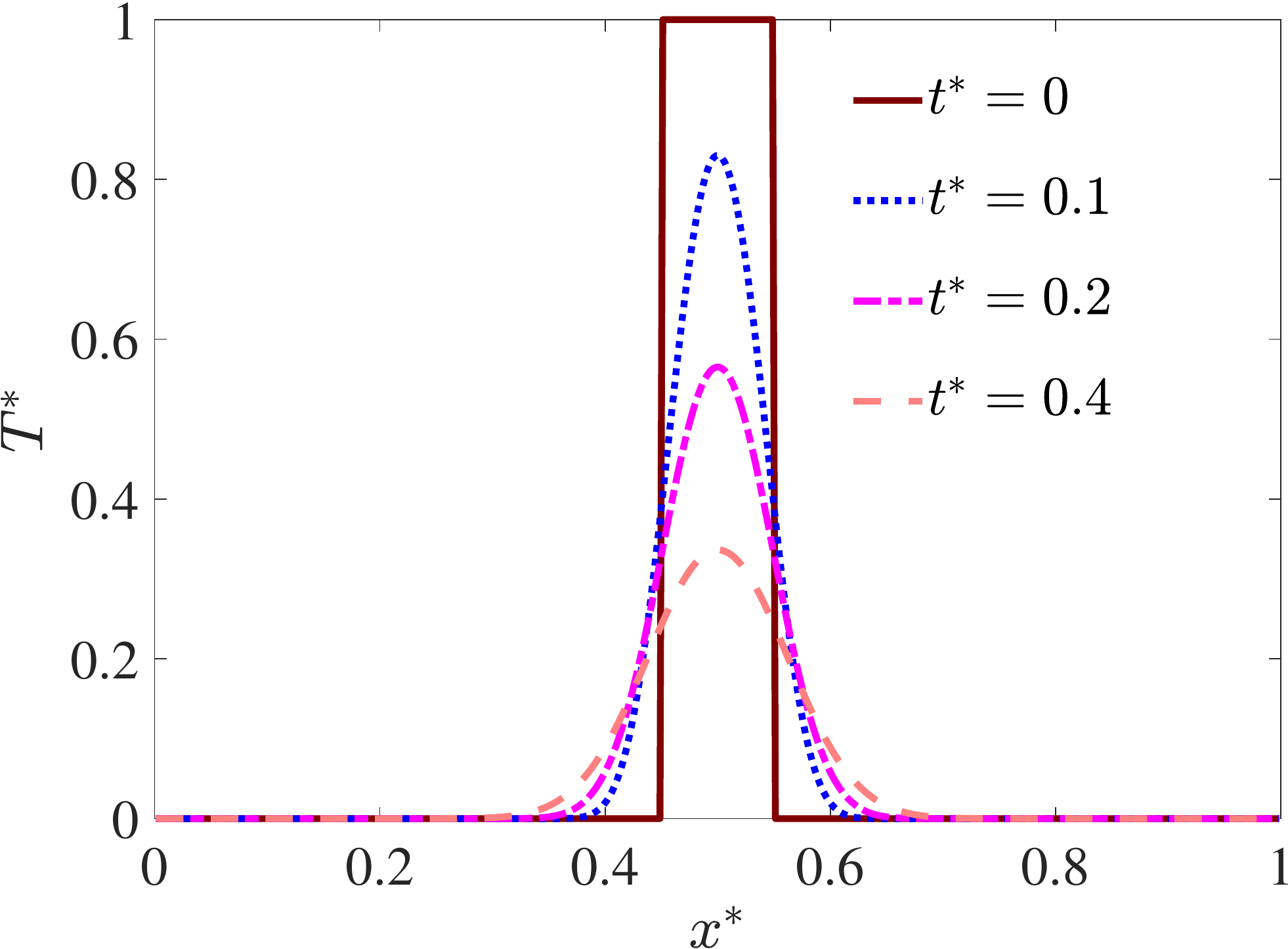}} \\
 \caption{The distributions of spatio-temporal temperature $T^*(x^*,t^*)$ along the radial direction in quasi-two dimensional simulations (\cref{problemdescription}(b)) based on dimensionless phonon BTE (Eqs.~\eqref{eq:dimensionlessBTE},~\eqref{eq:dimensionlessparameters}), where $T^*=(T-T_0)/(T_h -T_0)$, $t^* \in [0,0.45)$, $x^* \in [0,1]$. }
 \label{2DSSUN}
\end{figure}
\begin{table}[htb]
\caption{The negative dimensionless temperature signals ($T^* <0 $) in quasi-two dimensional transient heat propagation simulations (\cref{problemdescription}(b)), where `YES' (`NO') represents that the negative dimensionless temperature signals can (cannot) be predicted obviously. }
\centering
\begin{tabular}{|*{6}{c|}}
 \hline
 & $\text{Kn}_R = \infty$  &  $\text{Kn}_R = 10.0 $  &$\text{Kn}_R =  1.0 $  & $\text{Kn}_R = 0.1 $  &$\text{Kn}_R =0.01$      \\
 \hline
$\text{Kn}_N = \infty$ & NO & NO& NO& NO& NO \\
 \hline
 $\text{Kn}_N = 10.0  $ & NO & NO& NO& NO& NO \\
 \hline
 $\text{Kn}_N = 1.0 $ & NO & NO& NO & NO& NO \\
 \hline
 $\text{Kn}_N = 0.1  $ & YES & YES &YES & NO& NO \\
 \hline
 $\text{Kn}_N = 0.01 $ & YES & YES & YES & YES & NO \\
 \hline
\end{tabular}
\label{signal}
\end{table}

\subsection{Beyond quasi-one dimensional heat propagation}

When simulating the quasi-two dimensional transient heat propagation (\cref{problemdescription}(b)), all numerical settings are similar to those in quasi-one dimensional simulations except that $N_{cell}=400 \times 400$, $N_{\theta} =36$, $N_{\varphi}=18$.

\begin{figure}[htb]
 \centering
 \subfloat[$\text{Kn}_{R}= \infty $,$\text{Kn}_{N}= 0.1 $]{\includegraphics[scale=0.26,clip=true]{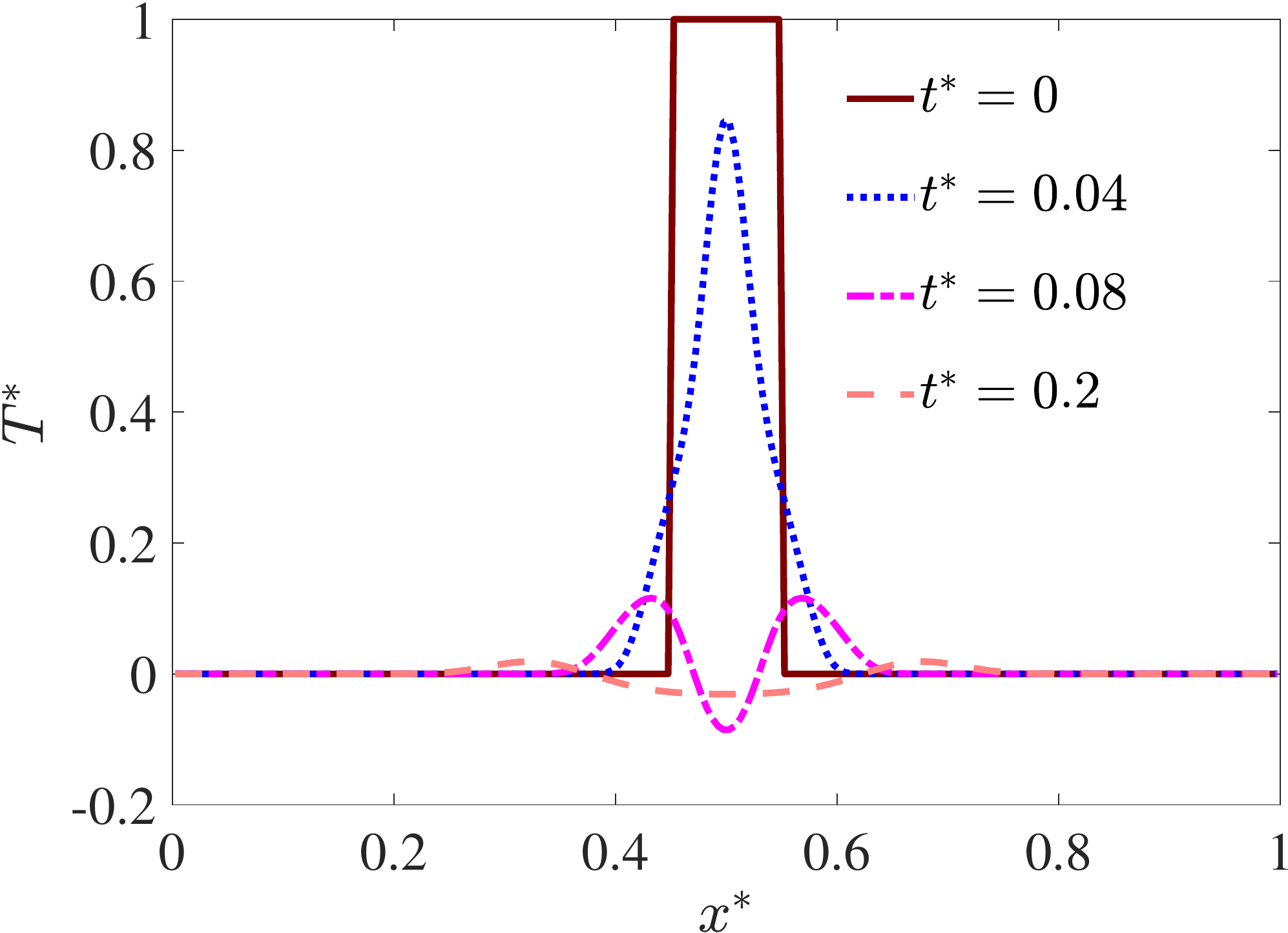} }~
  \subfloat[$\text{Kn}_{R}=1.0 $,$\text{Kn}_{N}=0.1 $]{\includegraphics[scale=0.26,clip=true]{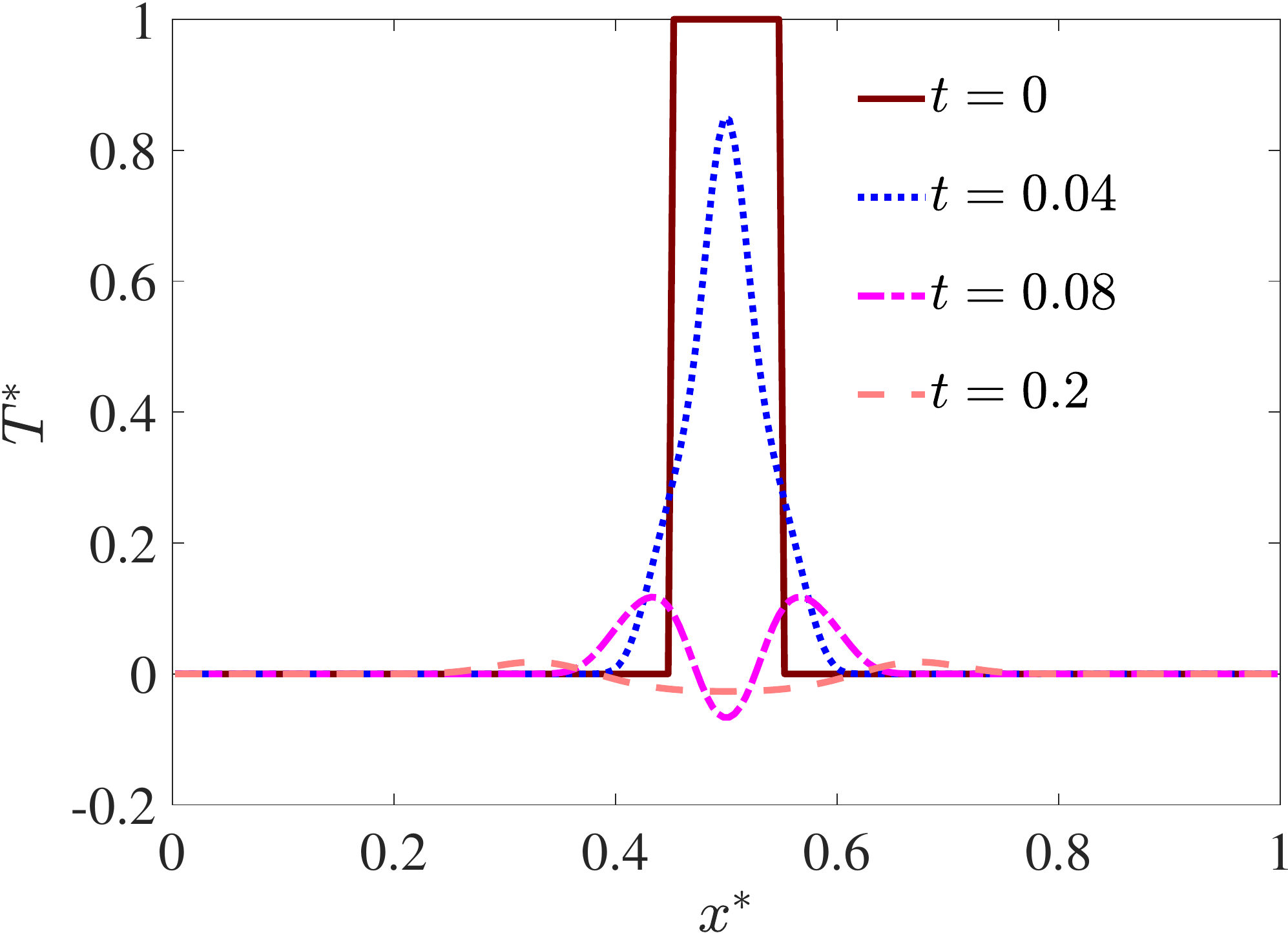} }\\
   \subfloat[$\text{Kn}_{R}=0.1 $,$\text{Kn}_{N}=1.0 $]{\includegraphics[scale=0.26,clip=true]{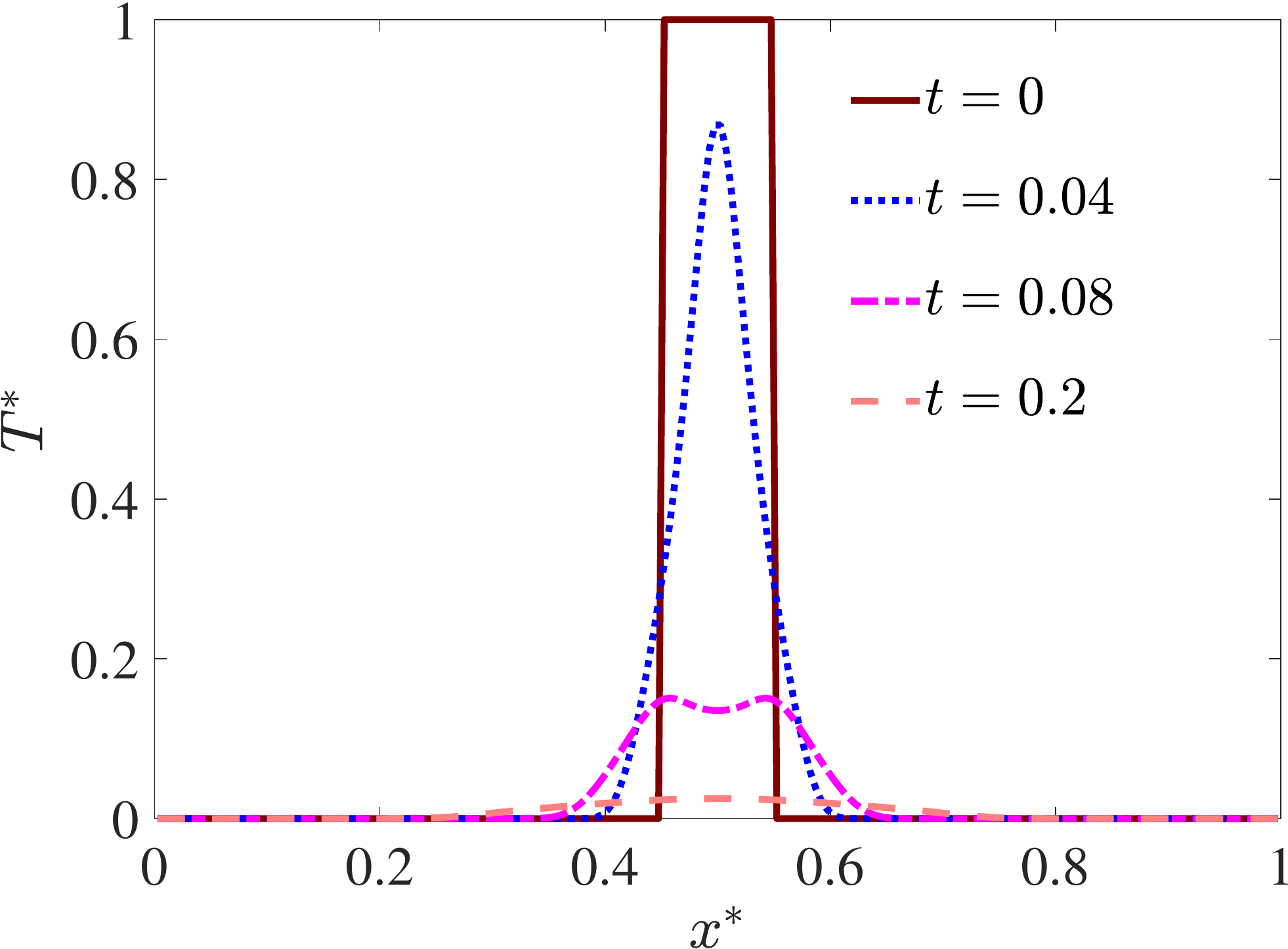} }~
    \subfloat[$\text{Kn}_{R}=0.1 $,$\text{Kn}_{N}=0.1 $]{\includegraphics[scale=0.26,clip=true]{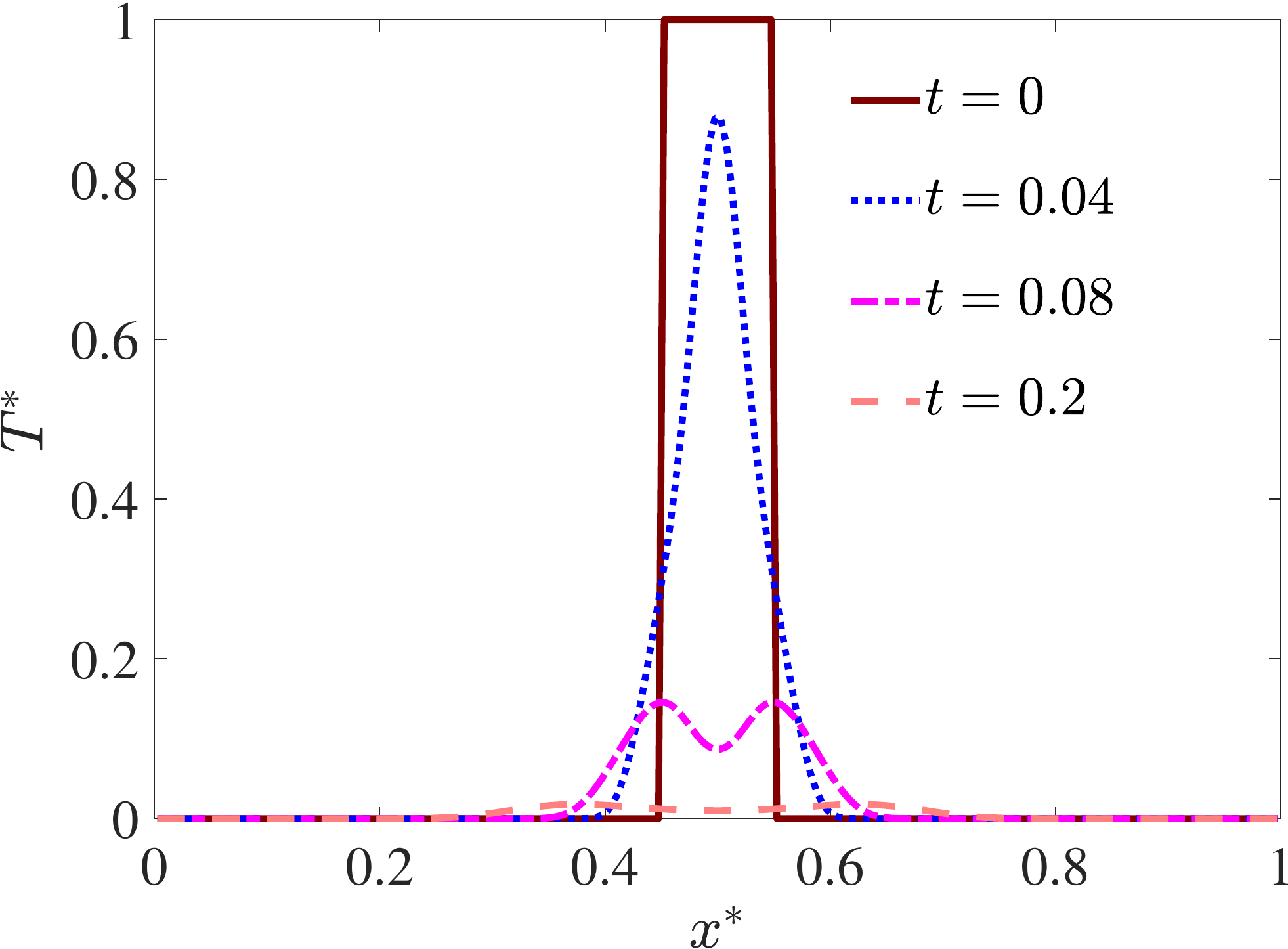} }\\
 \caption{The distributions of spatio-temporal temperature $T^*(x^*,t^*)$ along the radial direction in three dimensional system based on dimensionless phonon BTE (Eqs.~\eqref{eq:dimensionlessBTE},~\eqref{eq:dimensionlessparameters}), where $T^*=(T-T_0)/(T_h -T_0)$, $t^* \in [0,0.45)$, $x^* \in [0,1]$,  $N_{cell}=200 \times 200 \times 200$, $\text{CFL}=0.4$, $N_{\theta}= 24$, $N_{\varphi}=24$. }
 \label{3DSSUN}
\end{figure}

In the hydrodynamic regime, something amazing appears.
As shown in~\cref{2DSSUN}(c)(d), when the N scattering dominates heat conduction, $T^* < 0$ is satisfied in some areas when $t^*=0.2$ or $t^*=0.4$.
Namely, at sometime somewhere, the temperature $T(\bm{x},t)$ will be lower than the lowest environment temperature at the initial moment.
When the phonon scattering is not sufficient and the heat conduction is in the (quasi) ballistic regime, as shown in~\cref{2DSSUN}(a)(b)(e), this phenomenon disappears.
It also disappears when the resistive scattering dominates heat conduction.
{\color{black}{A number of numerical simulations are conducted with different $(\text{Kn}_R, \text{Kn}_N)$, as shown in Table.~\ref{signal}, where `YES' (`NO') represents that the negative dimensionless temperature signals $T^* <0 $ can (cannot) be predicted obviously.
It can be found that only as normal scattering dominates heat conduction, this novel transient phenomenon can appear.
The resistive scattering blocks the appearance of this negative signal.}}
Similar phenomenon is also observed in three dimensional heat conduction, as shown in~\cref{3DSSUN}.

{\color{black}{This negative signal $T^* < 0$ is caused by strong N scattering and the collective phonon drift in the hydrodynamic regime~\cite{PhysRev_GK,PhysRev.148.766,PhysRevX.10.011019}.
In this regime, $f \rightarrow  f_N^{eq}$ and the phonon BTE can recover the phonon wave equation or other hyperbolic equations (e.g., GK equation) within certain approximations according to the Chapman-Enskog expansion~\cite{PhysRev_GK,PhysRev.148.766,WangMr15application,PhysRevB_SECOND_SOUND,PhysRevX.10.011019}.
The analytical solutions~\cite{tikhonov2013equations} of heat wave equation~\eqref{eq:wavetemerpature} (see \ref{sec:anawave}) show that the transient temperature is totally determined by the initial or boundary conditions.
Within certain initial condition or settings, $T (x, t) <T_0$ is possible in quasi-2D or 3D systems.

Actually, when both $\text{Kn}_N$ and $\text{Kn}_R$ are finite and nonzero, the transient heat propagation is a competition result of the thermal dissipation and wave propagation.
When $\text{Kn}_N$ or $\text{Kn}_R$ is small, from a phenomenological point of view, it can be approximately described by such a hyperbolic heat conduction equation~\cite{PhysRevB_SECOND_SOUND,PhysRevX.10.011019,second_sound_ge2020,RevModPhysJoseph89},
\begin{align}
c_1 \frac{ \partial^2 T }{ \partial t^2 } + c_2  \frac{ \partial T }{ \partial t } + c_3 \nabla_{\bm{x}}^2 T = Q,
\label{eq:wavehyperbolic}
\end{align}
where $c_1$, $c_2$, $c_3$ are coefficients and $Q$ is the external heat source.
The combination of the first and third terms on the left side of Eq.~\eqref{eq:wavehyperbolic} control the heat wave.
While the combination of the second and third terms on the left side of Eq.~\eqref{eq:wavehyperbolic} control the thermal dissipation and temperature damping.
When R scattering dominates phonon transport, heat dissipation terms dominate heat transfer, the negative signal disappears.
When N scattering is very strong and R scattering is very weak, the heat wave terms plays a leading role on transient thermal transport, and the negative signal is possible according to the analytical solution of wave equation (see \ref{sec:anawave}).

In addition, according to previous studies~\cite{cepellotti_phonon_2015,lee2017,PhysRevB_SECOND_SOUND,PhysRevX.10.011019,RevModPhysJoseph89,shang_heat_2020}, the hydrodynamic heat propagation in low-dimensional materials (e.g., graphene~\cite{cepellotti_phonon_2015,shang_heat_2020}) or three-dimensional materials with nonlinear phonon dispersion (e.g., graphite~\cite{PhysRevX.10.011019}) can also be described by the phenomenological hyperbolic or wave equation.
In other words, maybe this negative signal, which is a characteristic of the wave equation, can also be predicted in these materials.
More detailed study on this topic can be done in the future.}}

To the best of our knowledge, this transient heat conduction phenomenon, which is like a drop of water hitting a horizontal surface and making a ripple, has never been mentioned in phonon areas.
Although the wave like propagation of heat has been observed in the quasi-one dimensional heat pulse experiments~\cite{ma2013a,PhysRevLett_secondNaF,Dreyer1993,kovacs2015,kovacs2018} or transient thermal grating (TTG) experiments~\cite{huberman_observation_2019,GuoZl16DUGKS,LUO2017970,collins_non-diffusive_2013} due to (quasi) ballistic or hydrodynamic phonon transport, the transient temperature is never smaller than the lowest temperature at the initial moment in these studies.
{\color{black}{Actually, compared to previous studies on hydrodynamic phonon transport~\cite{nie2020thermal,WangMr15application,PhysRevLett_secondNaF,Dreyer1993,kovacs2015,kovacs2018,huberman_observation_2019}, the main differences in this work are the initial conditions or numerical settings.
Fortunately, a novel transient hydrodynamic phenomenon was predicted.}}
According to present results and theoretical analysis, this phenomenon can only appear beyond quasi-one dimensional simulations.
In addition, it can only appear when the N scattering is strong and dominates heat conduction.
Based on this phenomenon, the hydrodynamic phonon transport can be well distinguished from the (quasi) ballistic or diffusive phonon transport.

\begin{figure}[htb]
 \centering
 \subfloat[] { \includegraphics[scale=0.5,viewport=50 50 450 390,clip=true]{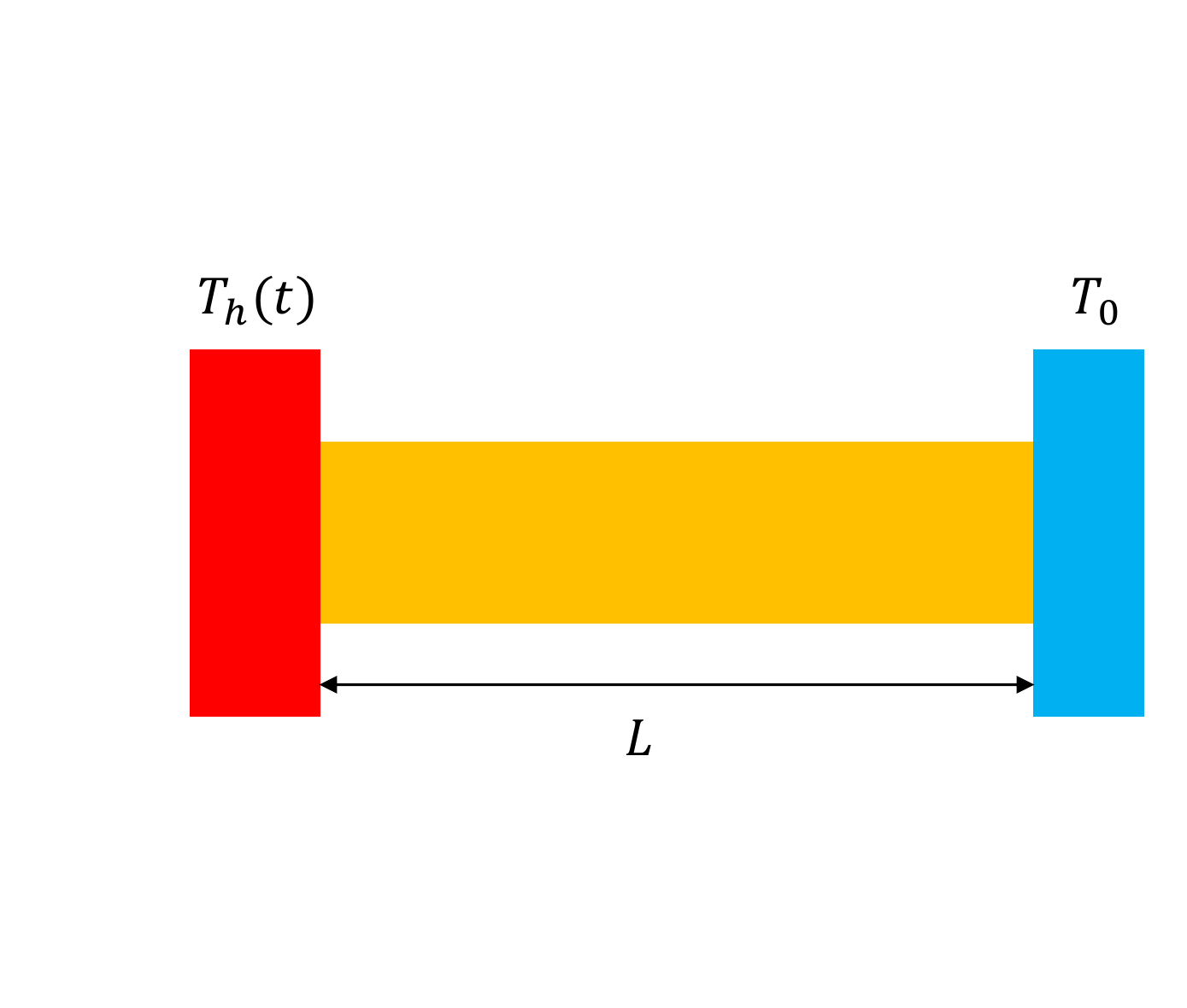} }
 \subfloat[] { \includegraphics[scale=0.5,viewport=500 50 800 390,clip=true]{period_12D.pdf} }
 \caption{Heat propagation with time varying heat source temperature. The temperature of heat source is $T_h (t)=T_0 +\Delta T/2+\Delta T/2 \cos(\omega_h t)$ and the temperature of heat sink is fixed at $T_0$, where $\omega_h =4 \pi v_g/L$. $T_h (t) \geq  T_0$ so that heat always flows from the heat source to the heat sink. The red area is heat source, orange area is the computational domain, blue area is the heat sink. At initial moment $t=0$, the temperature inside the system is $T_0$. (a) Quasi-one dimensional simulations~\cite{second_sound_ge2020}, the system length is $L$. (b) Quasi-two dimensional simulations with a heat source at the center. The diameters of the heat source and heat sink are $L$ and $L_h=L/10$, respectively. }
 \label{perioddescription}
\end{figure}
\begin{figure}[htb]
 \centering
 \subfloat[$\text{Kn}_{R}= \infty $,$\text{Kn}_{N}= 10.0 $]{\includegraphics[scale=0.26,clip=true]{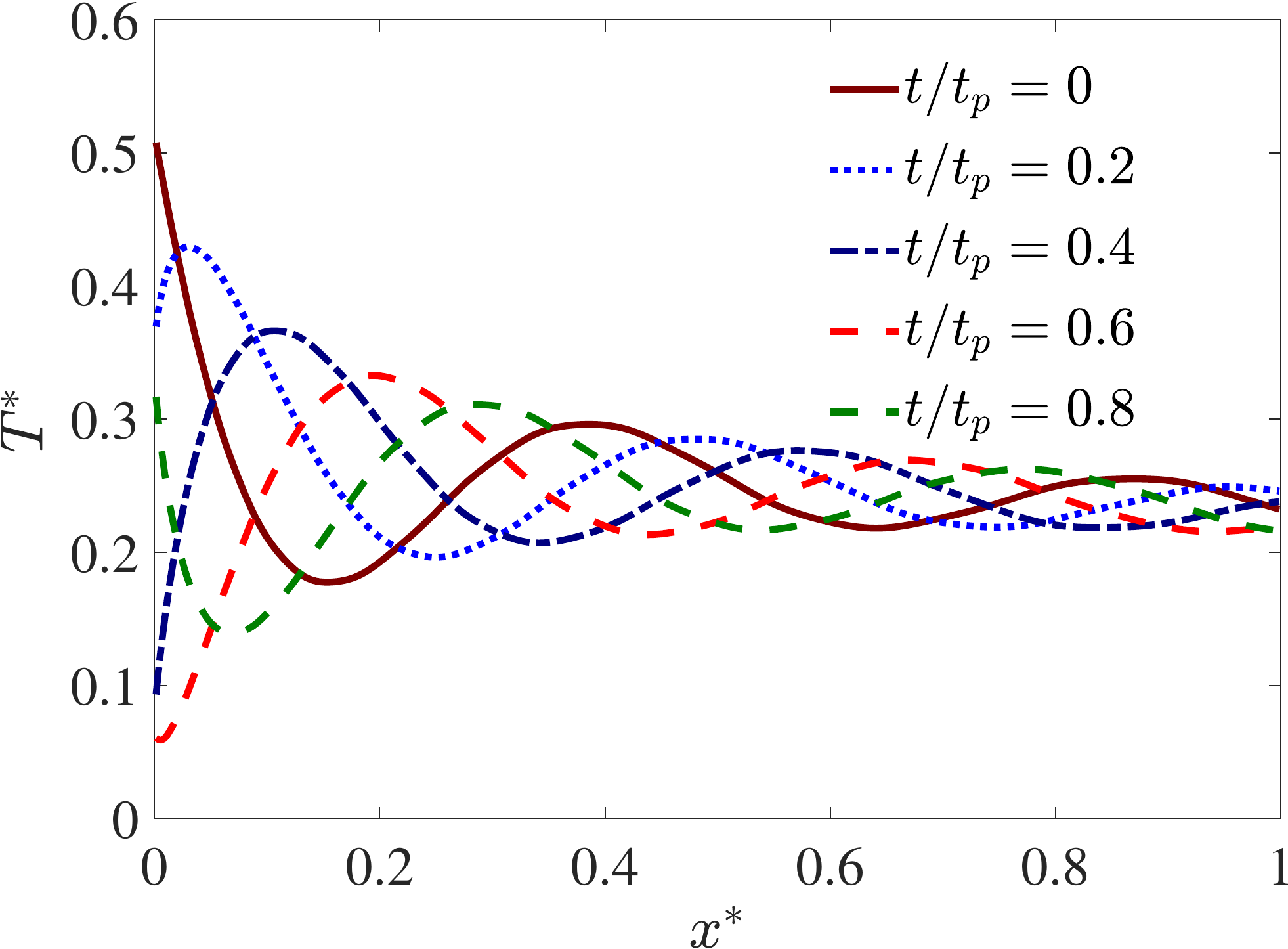}}~
 \subfloat[$\text{Kn}_{R}= \infty $,$\text{Kn}_{N}= 1.0 $]{\includegraphics[scale=0.26,clip=true]{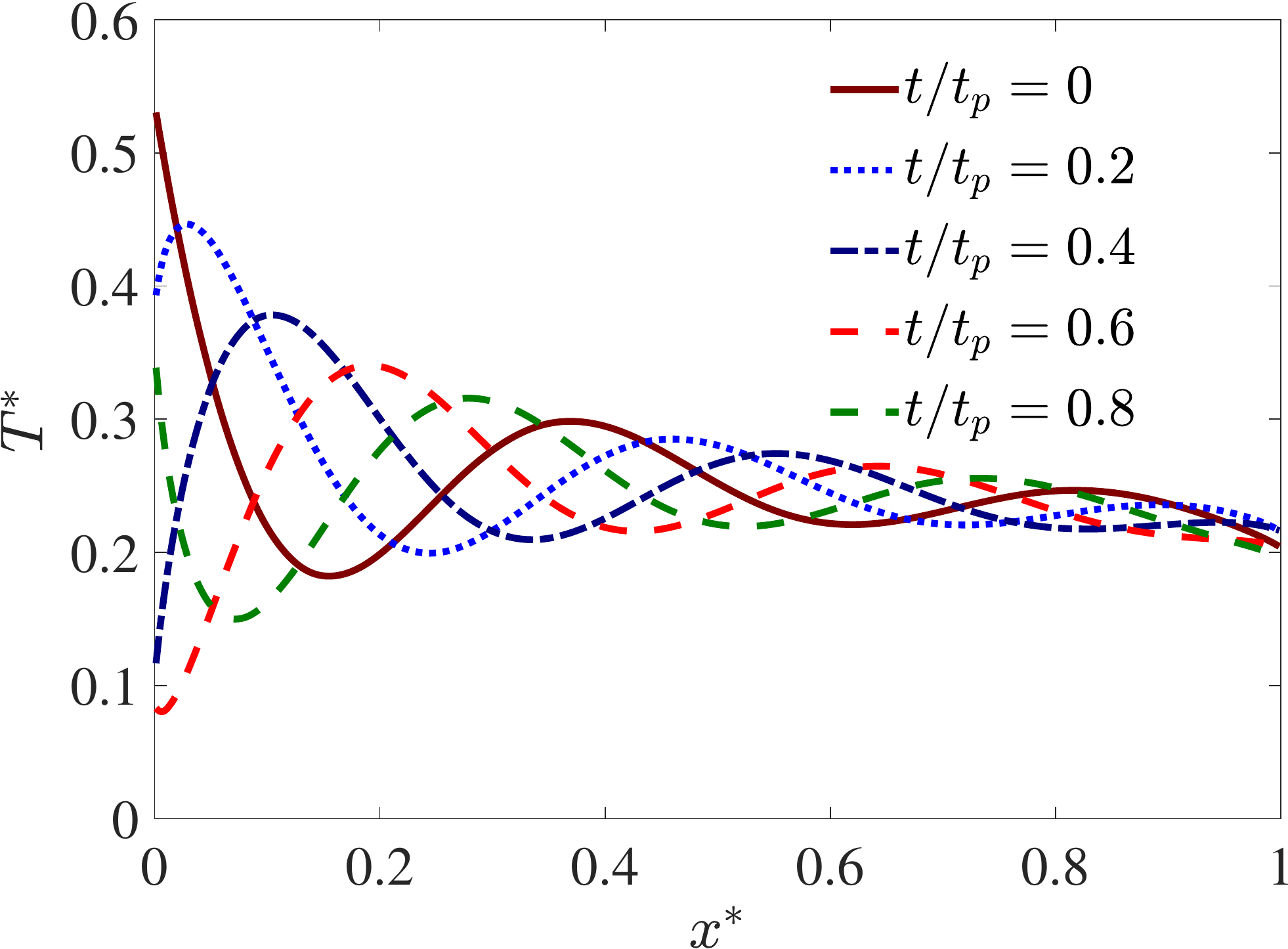}}~
 \subfloat[$\text{Kn}_{R}= \infty $,$\text{Kn}_{N}= 0.1 $]{\includegraphics[scale=0.26,clip=true]{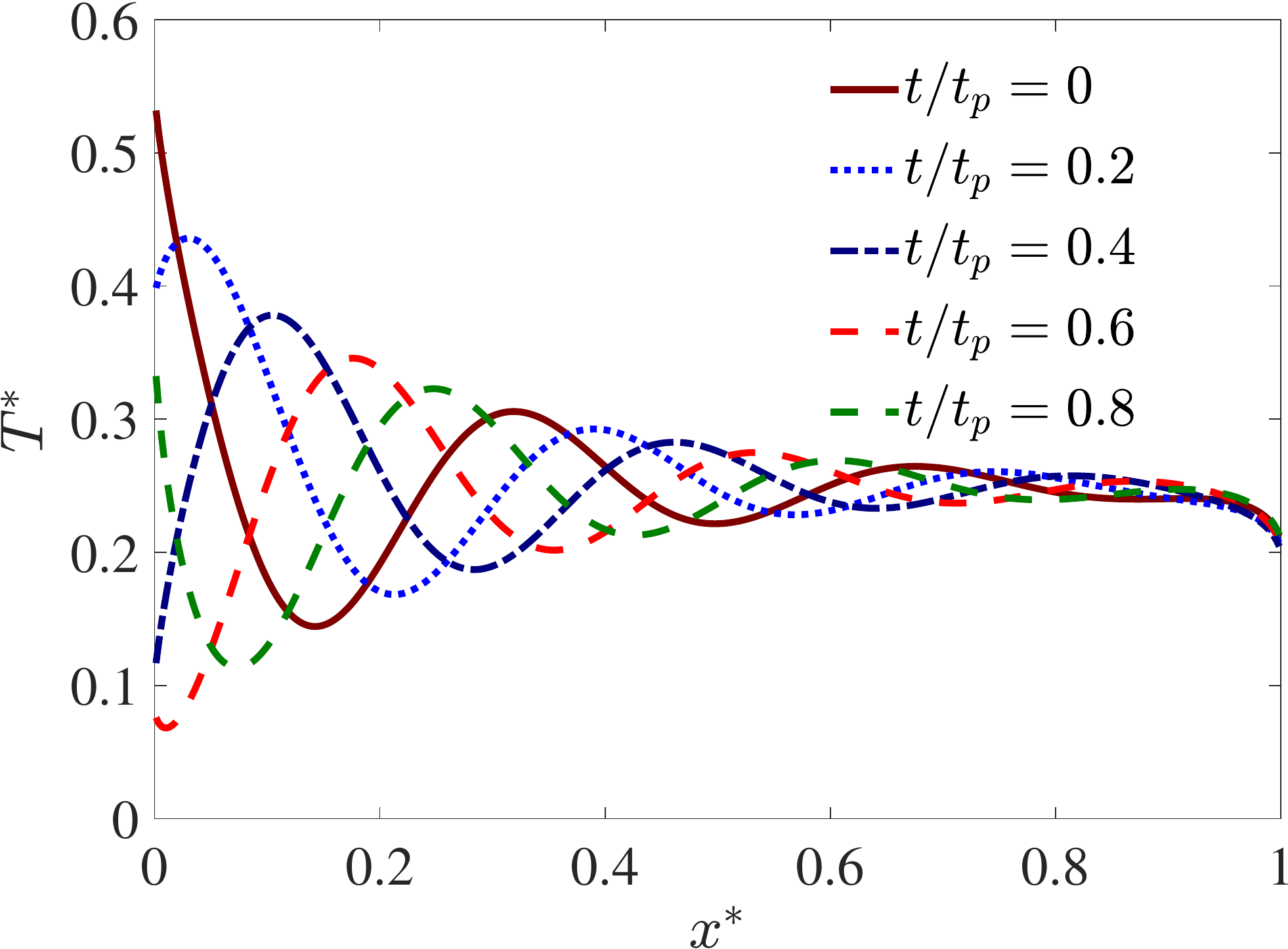}}\\
 \subfloat[$\text{Kn}_{R}= \infty $,$\text{Kn}_{N}= 0.01 $]{\includegraphics[scale=0.26,clip=true]{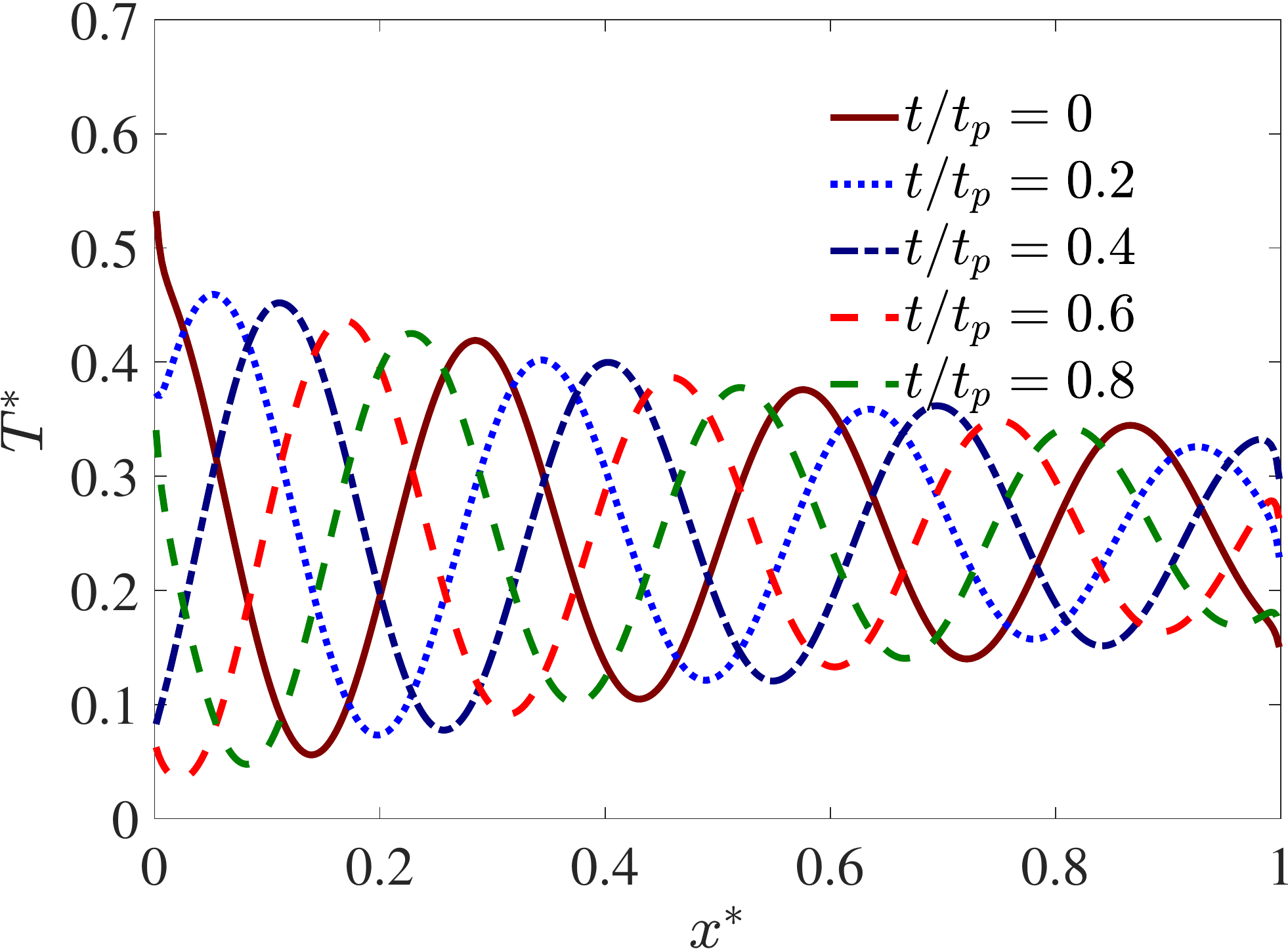}}~
 \subfloat[$\text{Kn}_{R}= 0.1 $,$\text{Kn}_{N}= \infty $]{\includegraphics[scale=0.26,clip=true]{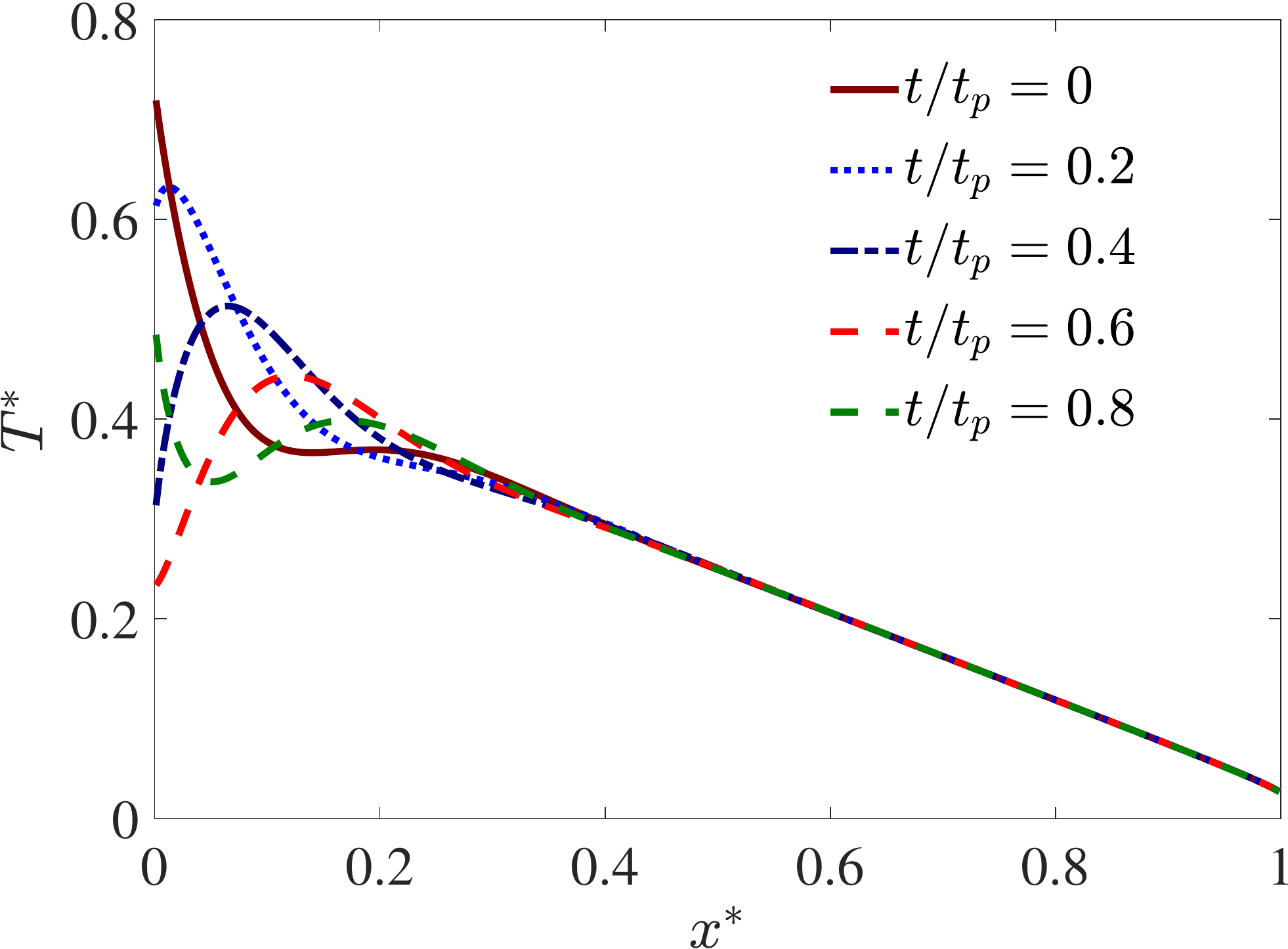}}~
 \subfloat[$\text{Kn}_{R}= 0.1  $,$\text{Kn}_{N}= 0.1 $]{\includegraphics[scale=0.26,clip=true]{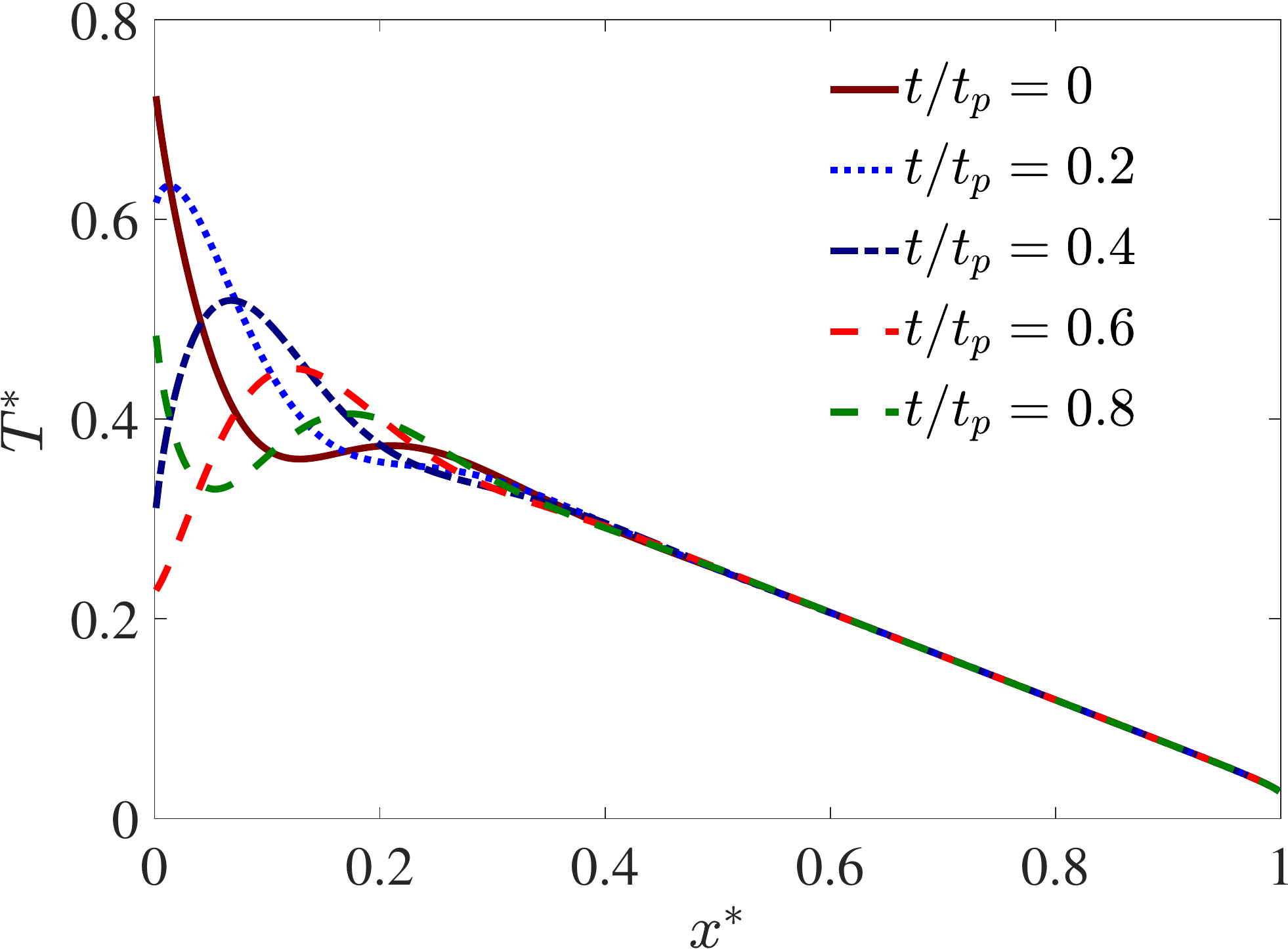}}  \\
 \caption{At periodic steady state, in quasi-one dimensional simulations (\cref{perioddescription}(a)), the distributions of the spatio-temporal temperature in one period $(t/t_p \in [0,1])$ with different $\text{Kn}_N$ and $\text{Kn}_R$, where $T^*=(T-T_0)/(T_h -T_0)$, $x^* =x/L \in (0.55,1]$.}
 \label{1dppUN}
\end{figure}
\begin{figure}[htb]
 \centering
 \subfloat[$\text{Kn}_{R}= \infty $,$\text{Kn}_{N}= \infty $]{\includegraphics[scale=0.26,clip=true]{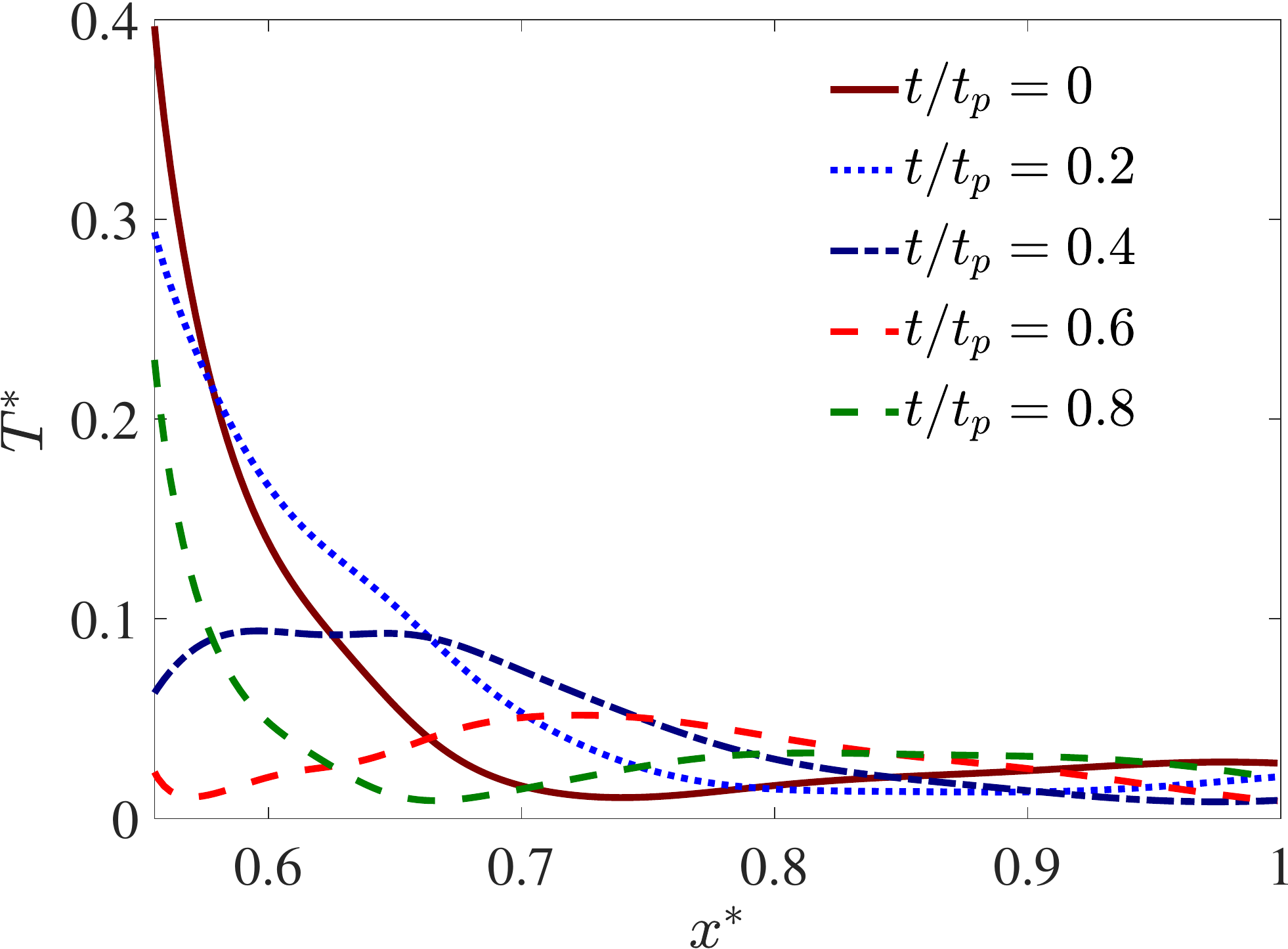}}~
 \subfloat[$\text{Kn}_{R}= \infty $,$\text{Kn}_{N}= 0.1 $]{\includegraphics[scale=0.26,clip=true]{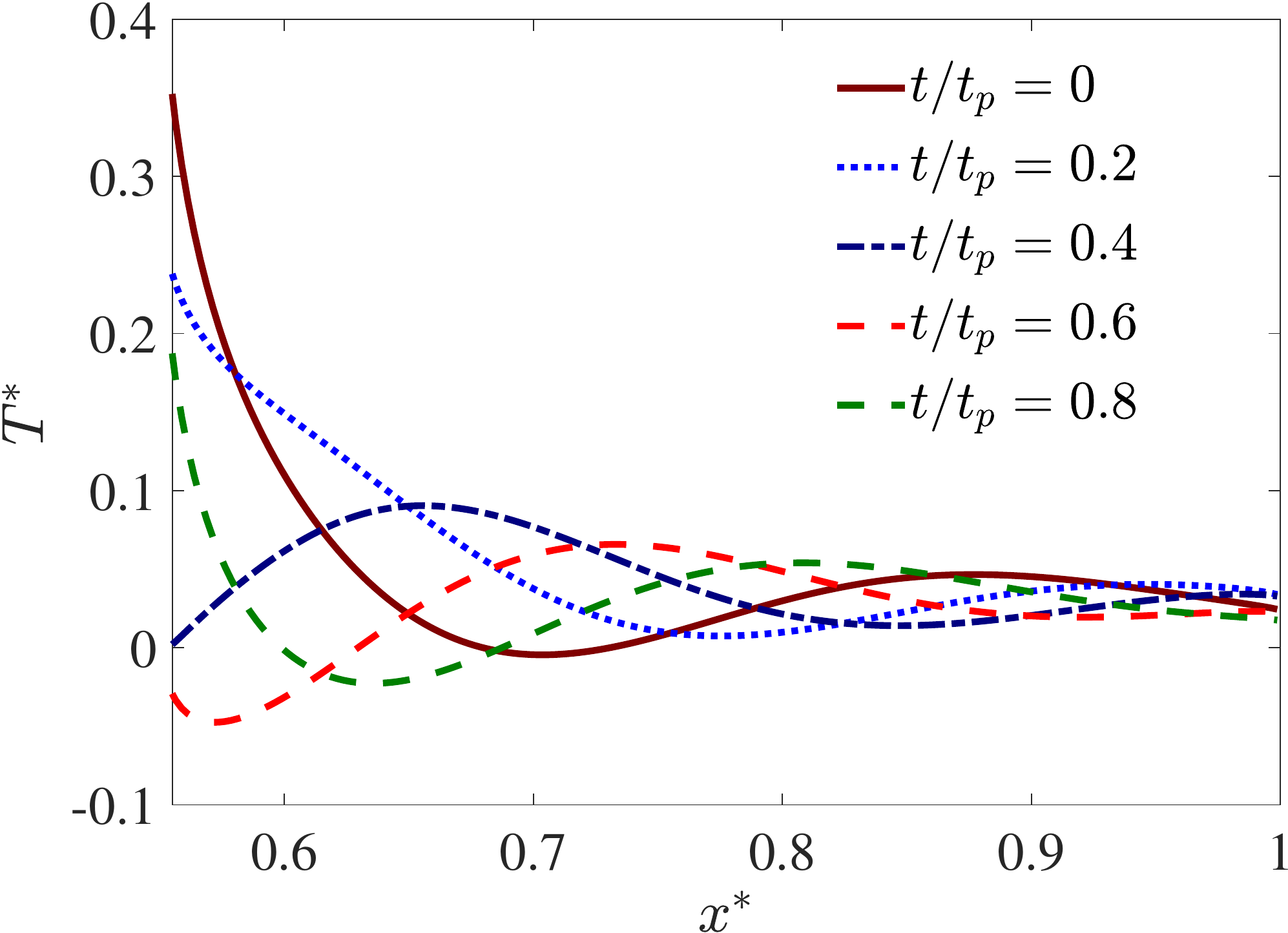}}~
 \subfloat[$\text{Kn}_{R}= \infty  $,$\text{Kn}_{N}= 0.01 $]{\includegraphics[scale=0.26,clip=true]{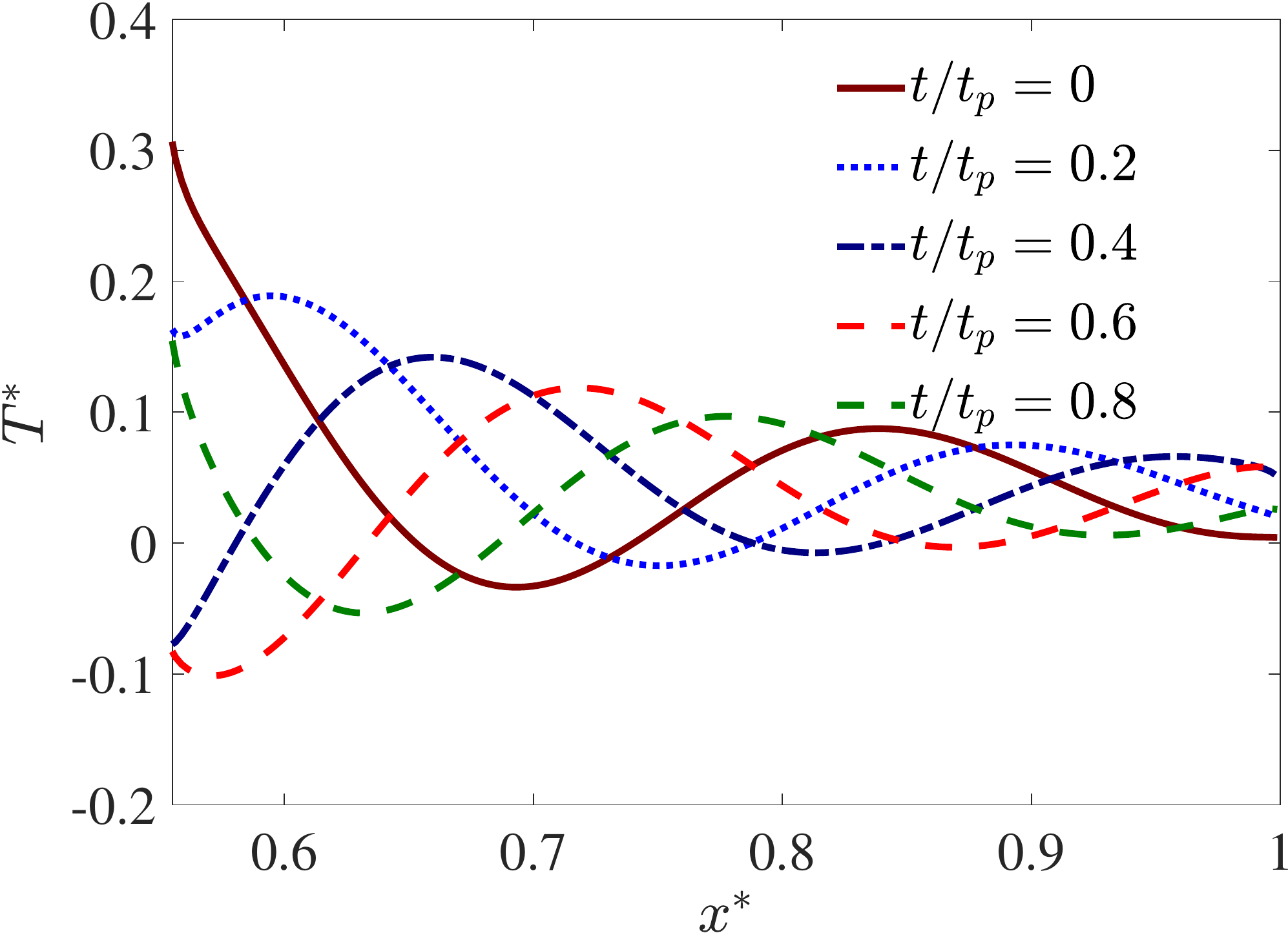}} \\
 \subfloat[$\text{Kn}_{R}= 1.0 $,$\text{Kn}_{N}= 10.0 $]{\includegraphics[scale=0.26,clip=true]{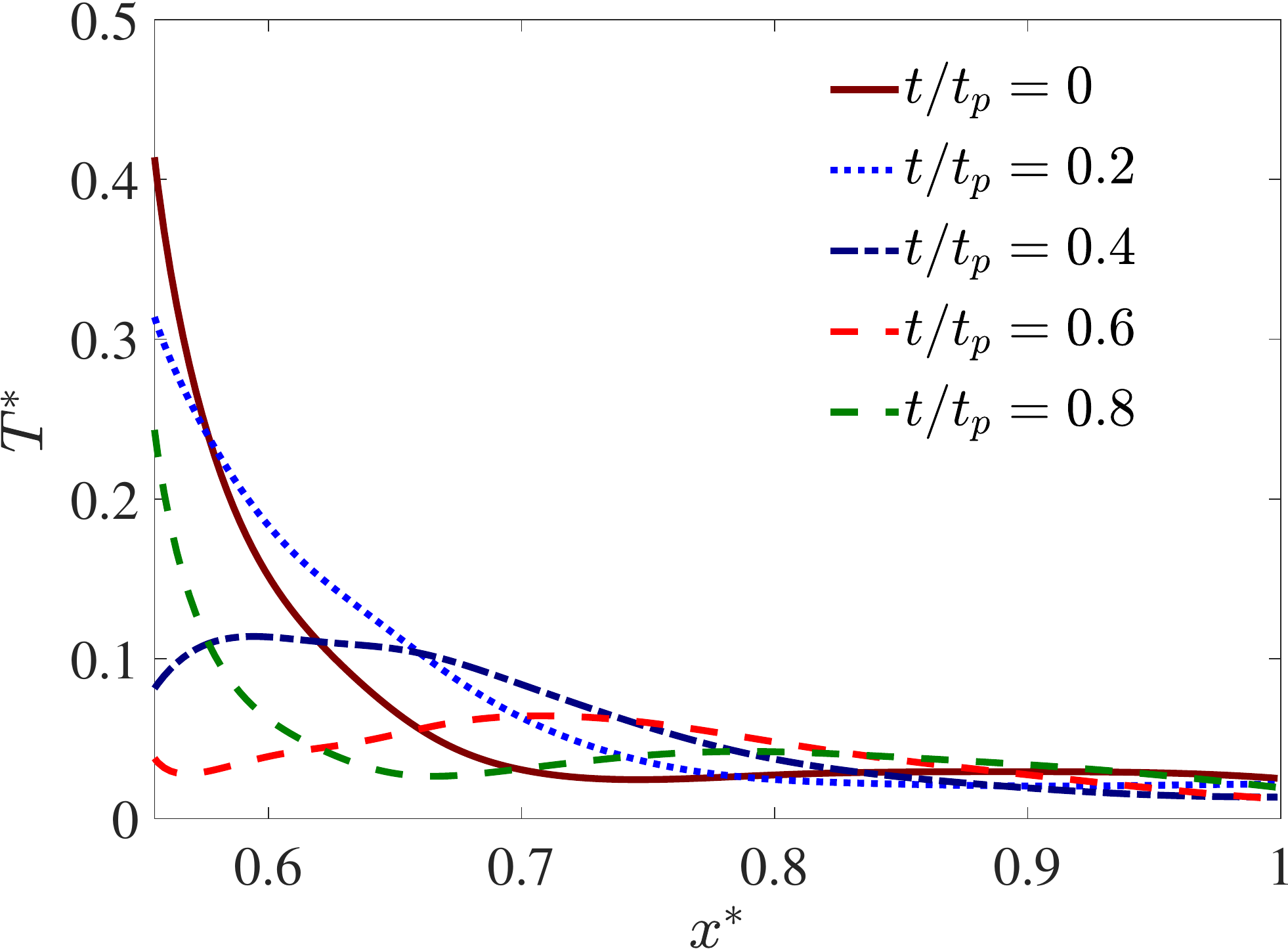}}~
 \subfloat[$\text{Kn}_{R}= 1.0 $,$\text{Kn}_{N}= 1.0 $]{\includegraphics[scale=0.26,clip=true]{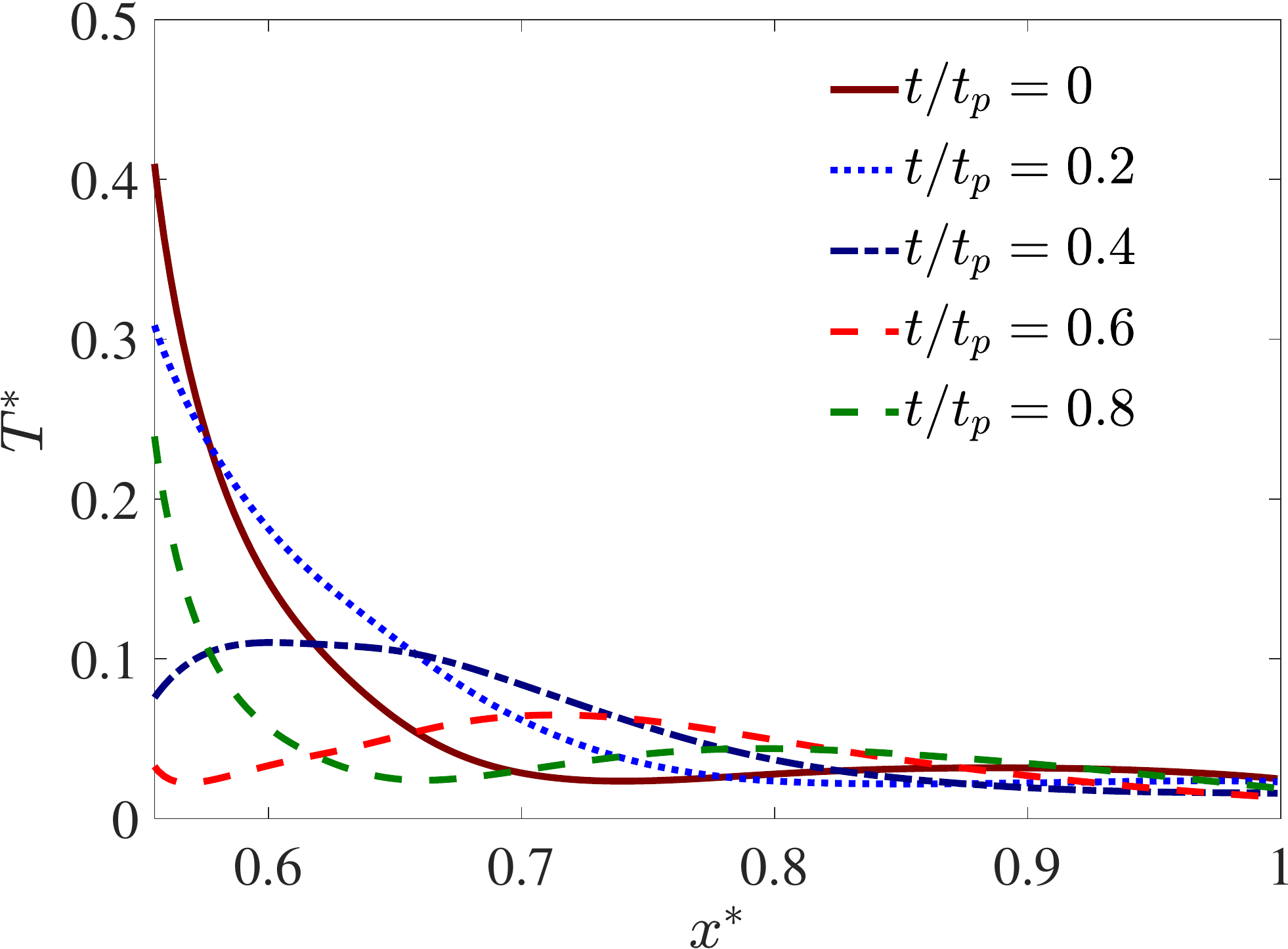}}~
 \subfloat[$\text{Kn}_{R}= 0.1  $,$\text{Kn}_{N}= 0.1 $]{\includegraphics[scale=0.26,clip=true]{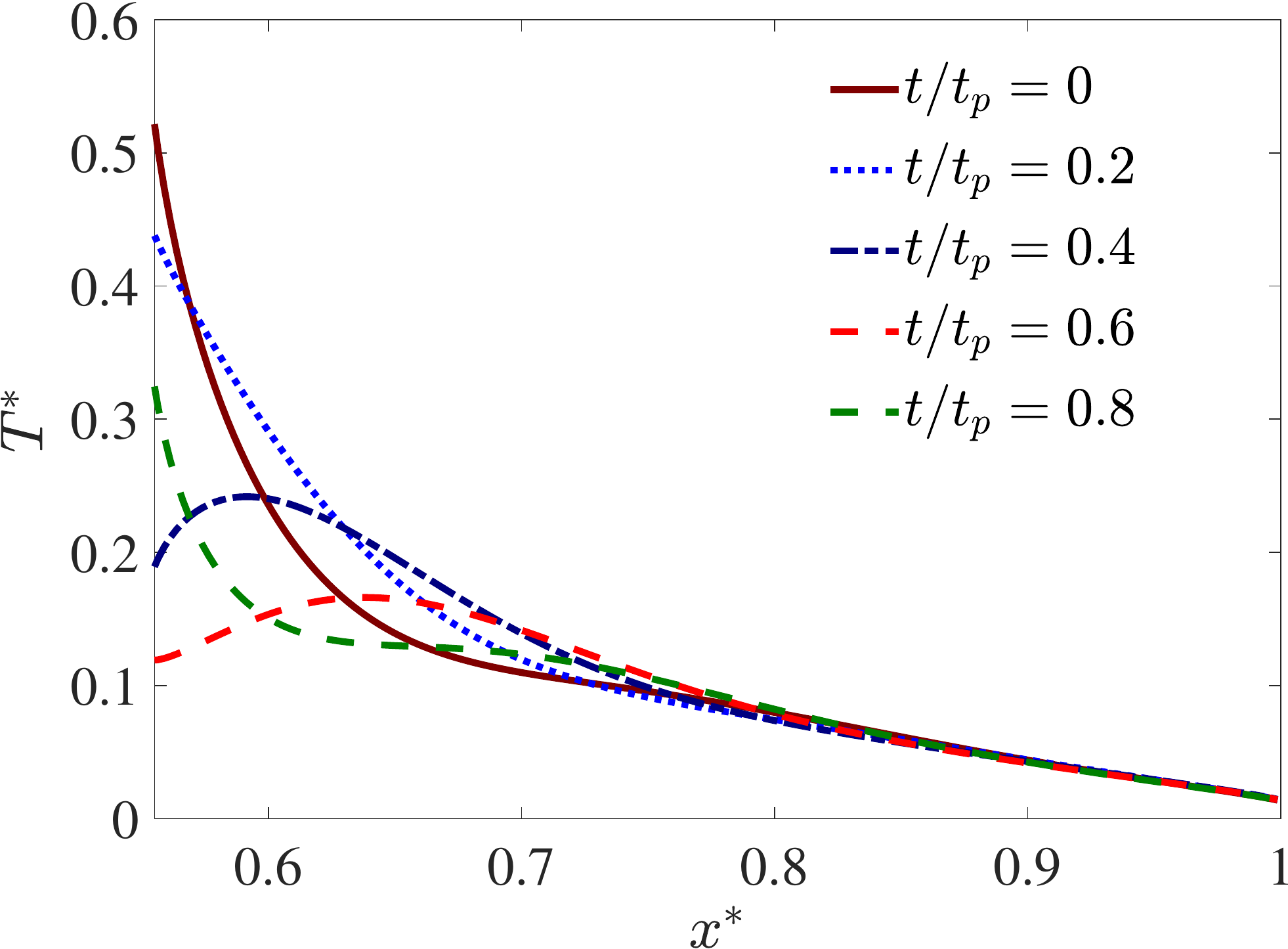}}  \\
 \caption{At periodic steady state, in quasi-two dimensional simulations (\cref{perioddescription}(b)), the distributions of the spatio-temporal temperature along the radial direction in one period $(t/t_p \in [0,1])$ with different $\text{Kn}_N$ and $\text{Kn}_R$, where $T^*=(T-T_0)/(T_h -T_0)$, $x^* =x /L \in (0.55,1]$. }
 \label{2dppUN}
\end{figure}
\section{Heat propagation with time varying heat source}
\label{sec:timevarying}

Note that in many transient thermal experiments~\cite{huberman_observation_2019,PhysRevB.101.075303,second_sound_ge2020,zobeiri2019}, the heat source is time-dependent, and the laser heating and detection are performed at the same spatial point.
In order to make the simulation results closer to the experimental settings, the heat propagation with time varying heat source is studied and we mainly focus on the temperature change near the heat source.

The schematic of the heat propagation problem is shown in~\cref{perioddescription}.
In quasi-one dimensional simulations~\cite{second_sound_ge2020}, the system length is $L$ coupled with a heat source and heat sink.
In quasi-two dimensional simulations, a heat source is added at the center, the diameters of the heat source and heat sink are $L$ and $L_h=L/10$, respectively.
The temperature of heat source is $T_h (t)=T_0 +\Delta T/2+\Delta T/2 \cos(\omega_h t)$ and the temperature of heat sink is fixed at $T_0$.
The period of heating frequency is $t_p =2 \pi/ \omega_h = L/(2 v_g)$, which is comparable to the system reference time $t_{\text{ref}}$.
At initial moment $t=0$, the temperature inside the system is $T_0$.

The DUGKS~\cite{guo_progress_DUGKS,GuoZl16DUGKS,LUO2017970,zhang_discrete_2019,luo2019} is used to solve the phonon BTE.
The numerical discretization of the solid angle and spatial spaces is the same as those mentioned in the last section~\ref{sec:quasionecase1} except $\text{CFL}=0.4$.
An error $\epsilon$ is introduced and defined as the deviations of the temperature field over the whole computational domain in two successive periods, i.e.,
\begin{align}
\epsilon = \frac{ \sum_{i=1}^{N_{cell}} (T (\bm{x}_i, t_p \times j +t_p ) -T (\bm{x}_i ,t_p \times j ))^2  }{  N_{cell} \Delta T^2   },
\end{align}
where $\bm{x}_i$ is the central position of the discretized cell $i$ and $j>0$ is an integer.
The system reaches periodic steady state as $\epsilon <10^{-6}$.

In this work, we only focus on when and where $T^*(\bm{x}^*,t^*) <0$ (i.e., $T(\bm{x},t) < T_0$) is satisfied.
The quasi-one dimensional numerical results are shown in~\cref{1dppUN}.
When R scattering is weak and $\text{Kn}_N$ decreases gradually, as shown in~\cref{1dppUN}(a)-(d), the wave like propagation of heat can appear in both the (quasi) ballistic and hydrodynamic regimes, and the amplitudes of temperature near the heat source for different $\text{Kn}_N$ are similar.
It indicates that it is not easy to distinguish the (quasi) ballistic or hydrodynamic phonon transport only by the wave like propagation of heat.
When R scattering is strong, as shown in~\cref{1dppUN}(e)(f), $T^* < 0$ is impossible.
In addition, $T^* <0$ is not satisfied in quasi-one dimensional simulations.

In quasi-two dimensional simulations, in the ballistic regime, no phonon scattering happens and according to above analysis~\ref{ballistic_diffusive}, $T^* <0$ is impossible, as shown in~\cref{2dppUN}(a).
From~\cref{2dppUN}(b)(c), it can be found that when the N scattering is much stronger than the R scattering, something interesting happens, namely, $T^* <0$ appears near the heat source.
However, when R scattering dominates heat conduction, as shown in~\cref{2dppUN}(d)(e)(f), it causes heat dissipations and this phenomenon disappears.

\section{Conclusion and Outlook}
\label{sec:conclusion}

In this work, the transient heat propagation in homogeneous thermal system with a hotspot or time varying heat source is studied based on the frequency-independent phonon BTE under the Callaway model.
An interesting phenomenon is predicted, namely, beyond quasi-one dimensional cases, the transient temperature will be lower than the lowest value of initial environment temperature in the hydrodynamic regime within a certain range of time and space.
This phenomenon disappears in the (quasi) ballistic or diffusive regime.
It disappears in quasi-one dimensional simulations, either.
This novel transient heat propagation phenomenon of hydrodynamic phonon transport distinguishes it well from the heat propagation in the ballistic regime.
{\color{black}{In the future, the effects of real low-dimensional or bulk materials properties~\cite{huberman_observation_2019,PhysRevLett_ssNaf,lee2017,nanoletterchengang_2018,PhysRevLett_Strontium_Titanate,lee_hydrodynamic_2015,cepellotti_phonon_2015} on this novel hydrodynamic phenomenon could be investigated.}}

Based on the present results, in order to well distinguish the wave like propagation of heat in the (quasi) ballistic or hydrodynamic regime, a quasi-one dimensional thermal propagation is not ensured and quasi-two dimensional experiments are necessary, which have also been mentioned in the summary part of Ref~\cite{kovacs2018}.
According to previous experiments~\cite{zobeiri2019}, this unique phenomenon in the hydrodynamic regime may be measured by the Raman/TDTR/FDTR probing techniques with time varying heat source in the future.

{\color{black}{The most exciting thing is that, soon after our main result was online (https://arxiv.org/abs/2104.10981), this novel transient hydrodynamic phenomenon was measured by another independent research group in graphite by Picosecond Laser Irradiation~\cite{transient_cooling_2021} (https://arxiv.org/abs/2104.12343).}}

\section*{Author statements}

Chuang Zhang implements the numerical simulations of multiscale phonon transport and writes the paper.
Zhaoli Guo is the corresponding author and provides useful academic communications and guides.

\section*{Conflict of interest}

No conflict of interest declared.

\section*{Acknowledgments}

This study was supported by the National Natural Science Foundation of China (51836003).
The authors acknowledge Songze Chen, Jing-Tao L\"u, Manyu Shang, yangyu Guo and Samuel C. Huberman for communications of hydrodynamic phonon transport, acknowledge Chengyun Hua for communications of quasiballistic phonon transport, acknowledge Rulei Guo and Albert Beardo Ricol for useful introductions and communications about Raman/TDTR/FDTR experiments.

\appendix

\section{Temperature wave equation in the phonon hydrodynamic regime}
\label{sec:temperature_wave}

When the N scattering dominates heat conduction and the R scattering is ignored, the phonon BTE~\eqref{eq:BTE} becomes
\begin{align}
\frac{\partial e}{\partial t }+ v_g  \bm{s} \cdot \nabla_{\bm{x}} e  = \frac{e^{eq}_{N} -e}{\tau_{N}}.
\end{align}
Due to the energy and momentum conservation of the N scattering, taking zeroth- and first- orders moments of above equation leading to~\cite{PhysRev_GK,PhysRev.148.766,lee_hydrodynamic_2015,Nanalytical,nie2020thermal,shang_heat_2020}
\begin{align}
\frac{\partial E}{\partial t }+  \nabla_{\bm{x}}  \cdot  \bm{q}  & =0, \label{eq:zeroBTE} \\
\frac{\partial \bm{q} }{\partial t }+  \nabla_{\bm{x}}  \cdot  \bm{Q}  & =0, \label{eq:firstBTE}
\end{align}
where
\begin{align}
\bm{Q} = \int v_g^2 \bm{s} \bm{s} e d\Omega.
\end{align}
Combining Eqs.~\eqref{eq:zeroBTE} and~\eqref{eq:firstBTE} leading to
\begin{align}
\frac{\partial^2 E}{\partial t^2 } = \nabla_{\bm{x}}  \cdot  \nabla_{\bm{x}}  \cdot  \bm{Q} .
\end{align}
Based on previous studies~\cite{lee_hydrodynamic_2015,wangmr17callaway,shang_heat_2020}, above equation can be transformed into a wave equation of temperature by assuming $e = e^{eq}_N$, i.e.,
\begin{align}
\frac{\partial^2 T}{\partial t^2 } = \frac{1}{v_{ss} ^2}    \nabla^2_{\bm{x}} T ,
\end{align}
where $v_{ss}=v_g/ \sqrt{3}$.
Please note that this assumption $e = e^{eq}_N$ is usually not satisfied because $\tau_N \neq 0$ in most thermal systems.

\section{Analytical solutions of the temperature wave equation}
\label{sec:anawave}

The analytical solutions of the temperature wave equation~\eqref{eq:wavetemerpature} can be found in textbook~\cite{tikhonov2013equations}.
Here we take the transient heat propagation with an initial hotspot (Sec.~\ref{sec:hotspot}) as an example.
Its mathematical expression is
\begin{align}
\frac{\partial^2 T}{\partial t^2 } &= \frac{1}{v_{ss} ^2}    \nabla^2_{\bm{x}} T, \quad   r \geq 0, t> 0, \\
T(\bm{x},t) & = F ( r  ), \quad   t=0,  \\
\frac{ \partial T(\bm{x},t) }{\partial t} &=0, \quad  t=0 ,
\end{align}
where $r=\left\| \bm{x}- \bm{x}_0 \right\|_2 $ is the distance from the center of the hotspot $\bm{x}_0$, $F(r)$ is the initial temperature distributions $T(r,0)$ (\cref{problemdescription}),
\begin{align}
\begin{split}
F ( r )&= \left \{
\begin{array}{ll}
    T_h ,                    &  r  \leq  L_h/2, \\
    T_0 ,     & r > L_h/2 .
\end{array}
\right.
\end{split}
\label{eq:initialF}
\end{align}

For quasi-one dimensional temperature wave problem, the spatio-temporal distributions of temperature can be obtained according to D 'Alembert's formula~\cite{tikhonov2013equations}, i.e.,
\begin{align}
T( \bm{x} ,t) = \frac{ F( \left\| \bm{x}- \bm{x}_0 \right\|_2  - t/v_{ss} )   + F(\left\| \bm{x}- \bm{x}_0 \right\|_2  + t/v_{ss} )}{2} .
\label{eq:wave1D}
\end{align}
For quasi-two dimensional temperature wave equation, the analytical solution is~\cite{tikhonov2013equations}
\begin{align}
T( \bm{x}  ,t) = \frac{1}{2 \pi v_{ss} } \frac{\partial}{ \partial t} \iint_{D''}  \frac{F( \left\| \bm{x}''-\bm{x}_0 \right\|_2 )  }{ \sqrt{ (v_{ss} t)^2 - (\left\| \bm{x}''-\bm{x} \right\|_2 ) ^2   } } d \sigma,
\label{eq:wave2D}
\end{align}
where $d \sigma$ is the integral over the circular region $D''$ with radius $v_{ss} t $ and circular center $\bm{x}$. $\bm{x}''$ is the position of a point in this region $D''$.
In three dimensional system, the solution can be obtained according to Poisson's formula~\cite{tikhonov2013equations}, i.e.,
\begin{align}
T( \bm{x}  ,t) = \frac{1}{4 \pi v_{ss} } \frac{\partial}{ \partial t} \iint_{S'}  \frac{F( \left\| \bm{x}'-\bm{x}_0 \right\|_2  )  }{v_{ss} t} dS,
\label{eq:wave3D}
\end{align}
where $dS$ is the integral over the spherical surface $S'$ with radius $v_{ss} t $ and circular center $\bm{x}$.
$\bm{x}' $ is the position of a point on the surface $S'$.

As $r > L_h$ and $r-L_h < v_{ss} t <r+L_h$, based on Eq.~\eqref{eq:wave3D}, we can get
\begin{align}
T(r ,t)= T_0 + \frac{ r -v_{ss} t }{ 2 r} (T_h -T_0).
\end{align}
Obviously, $T < T_0$ could be satisfied within certain spatial position $r$ and time $t$ in three dimensional system.
Similar results can be obtained in quasi-two dimensional system.
However, based on Eqs.~\eqref{eq:wave1D} and~\eqref{eq:initialF}, $T < T_0$ is impossible in quasi-one dimensional system.

{\color{black}{Please note that this negative signal $T <T_0 $ is a characteristic of the wave equation.
As long as the transient heat propagation can be described by the wave equation, this phenomenon is possible with suitable initial or boundary conditions, which is not limited by the materials properties.}}

\section{Numerical validation of the DUGKS}
\label{sec:dugks_validation}

\begin{figure}[htb]
 \centering
 \subfloat[$\text{Kn}_{R}= \infty$,  $\text{Kn}_{N}=\infty $] { \includegraphics[scale=0.28,clip=true]{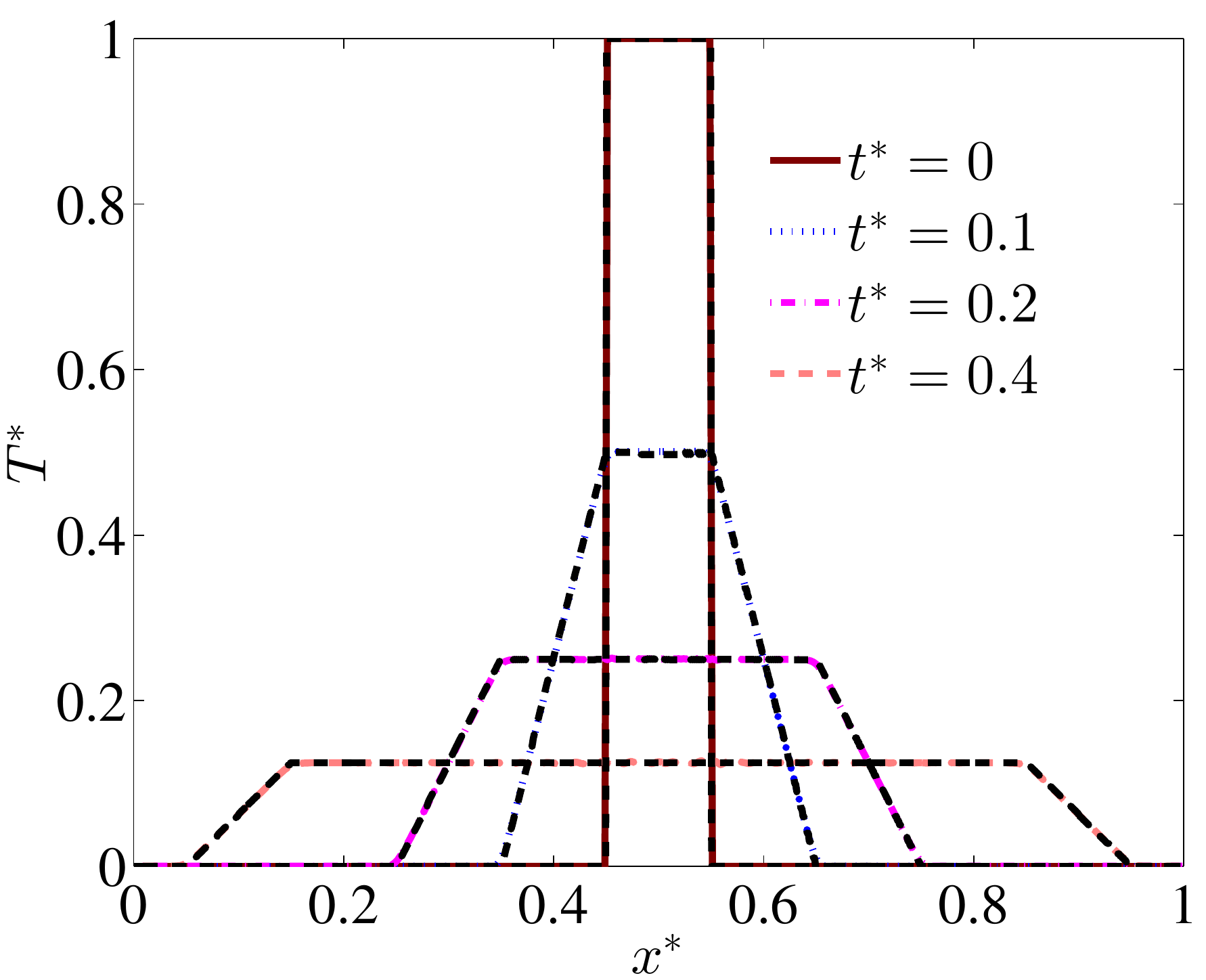} } ~
 \subfloat[$\text{Kn}_{R}= \infty $,  $\text{Kn}_{N}=0.001$] { \includegraphics[scale=0.28,clip=true]{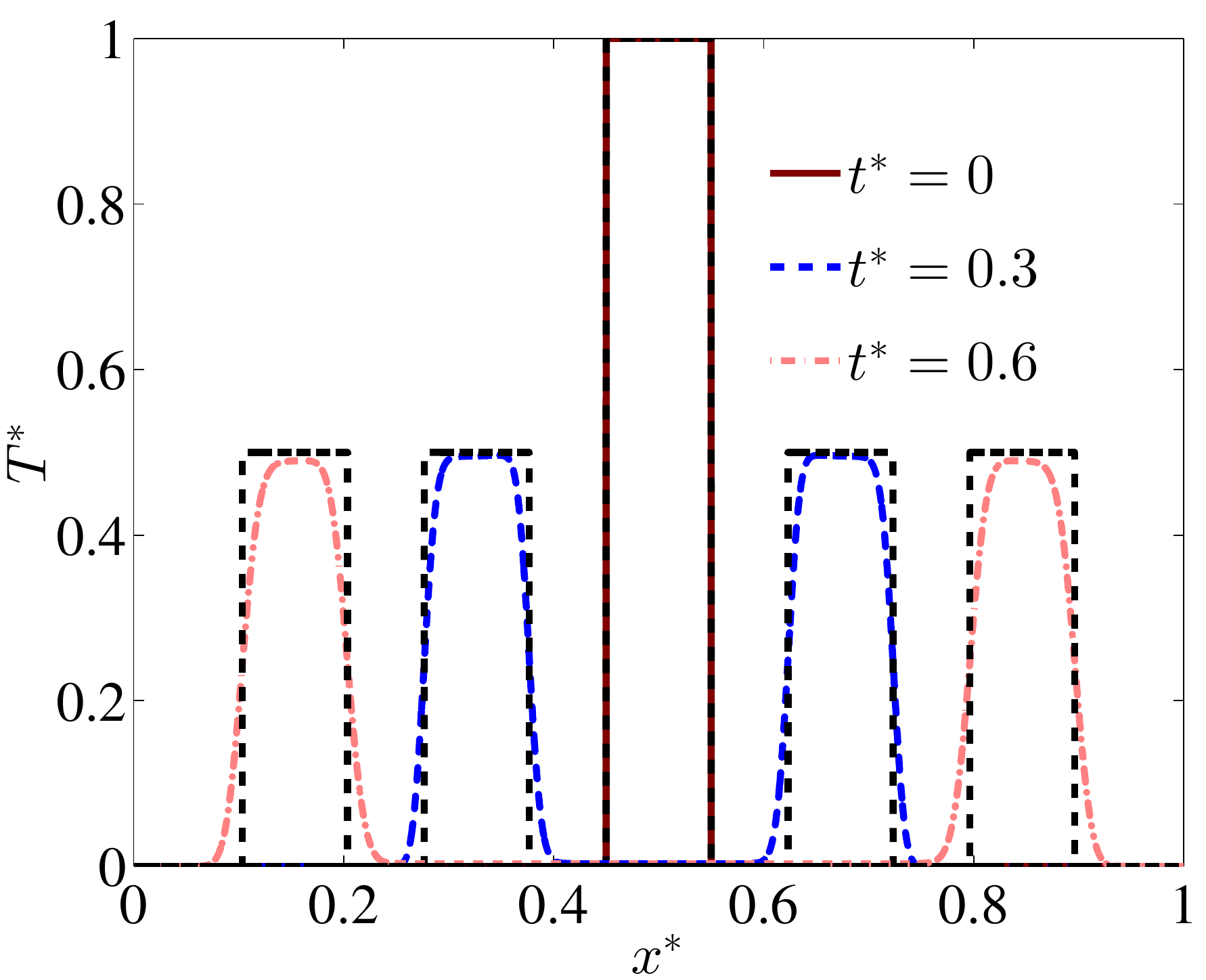} } ~
 \subfloat[$\text{Kn}_{R}= 0.01 $,  $\text{Kn}_{N}=\infty $] { \includegraphics[scale=0.28,clip=true]{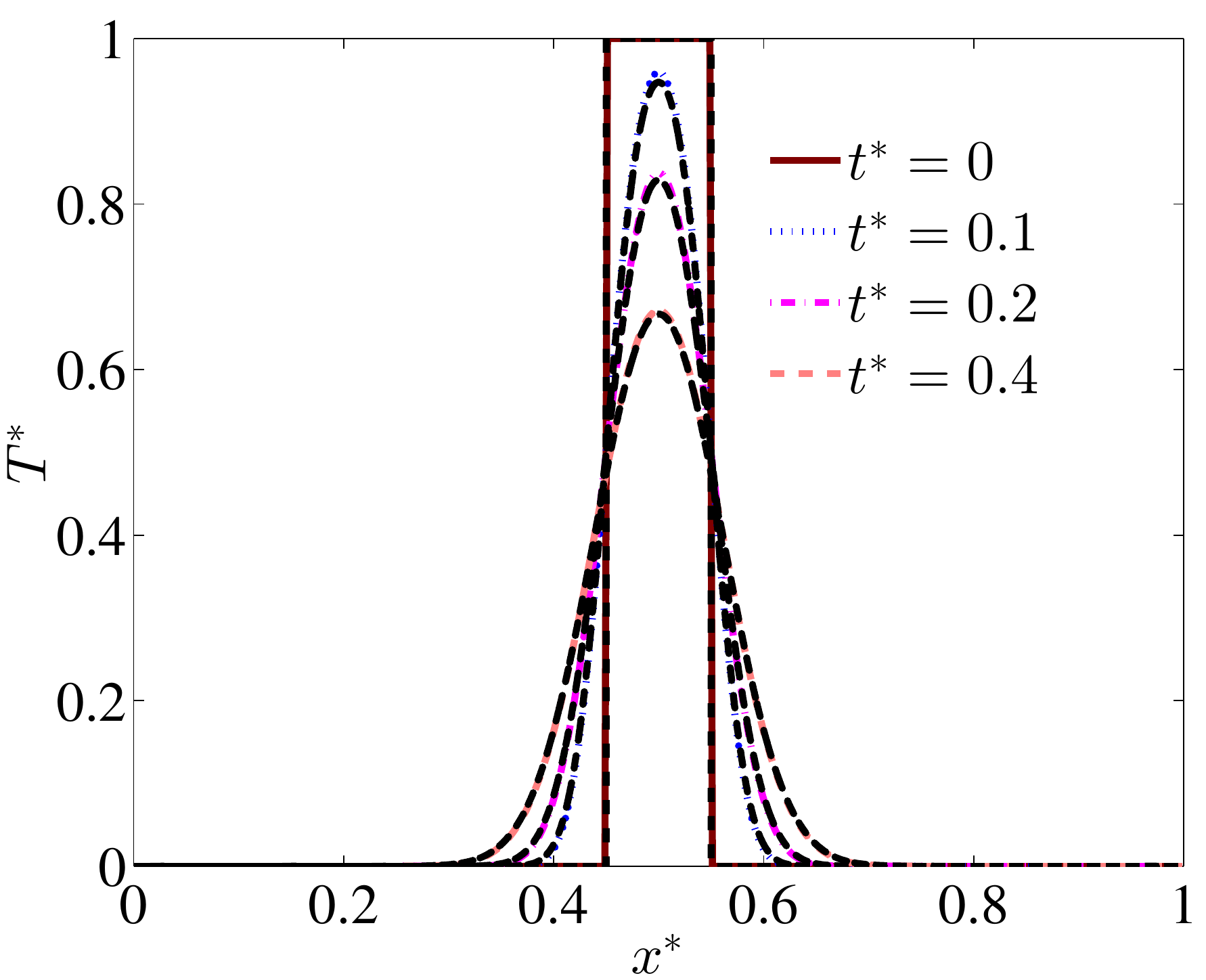} }   \\
 \caption{The distributions of spatio-temporal temperature $T^*(x^*,t^*)$ in quasi-one dimensional simulations (\cref{problemdescription}(a)) based on dimensionless phonon BTE (Eqs.~\eqref{eq:dimensionlessBTE},~\eqref{eq:dimensionlessparameters}), where $T^*=(T-T_0)/(T_h -T_0)$, $t^* \in [0,0.45)$, $x^* \in [0,1]$. Black dashed line is the analytical solutions of the phonon BTE in three phonon transport regimes. The dashed line are the results of DUGKS~\cite{guo_progress_DUGKS,GuoZl16DUGKS,LUO2017970,zhang_discrete_2019,luo2019}.
 (a) Ballistic limit (Eq.\eqref{eq:Tballistic}). (b) Hydrodynamic regime (Eq.~\eqref{eq:wave1D}). (c) Diffusive regime.  }
 \label{1Dthreelimits}
\end{figure}

A simple quasi-one dimensional numerical test of the DUGKS with an initial hotspot (Sec.~\ref{sec:hotspot}) is conducted.
The numerical settings are the same as those mentioned in section~\ref{sec:quasionecase1}.
From~\cref{1Dthreelimits}, it can be found that the numerical results predicted by DUGKS are in excellent agreement with the analytical solutions in the diffusive, ballistic and hydrodynamic regimes.
Therefore, the DUGKS is capable to describe the transient heat propagation in different regimes~\cite{guo_progress_DUGKS,GuoZl16DUGKS,LUO2017970,zhang_discrete_2019,luo2019}.

\section*{References}
\bibliographystyle{IEEEtr}
\bibliography{phonon}

\end{document}